\documentclass[aps,prl,twocolumn,a4paper,10pt,notitlepage,footinbib,superscriptaddress,showpacs]{revtex4-1}
\usepackage[english]{babel}
\usepackage[T1]{fontenc}
\usepackage{amssymb}
\usepackage{amsmath}
\usepackage{amsfonts}
\usepackage{textcomp}
\usepackage{graphicx,color}
\usepackage[dvipsnames]{xcolor}
\usepackage[utf8x]{inputenc}
\usepackage{mathtools}
\usepackage{float}

\usepackage{tabularx}

\begin{document}

\title{Emergence of effective temperatures in an out-of-equilibrium model of biopolymer folding} 

\vspace{-0.2 cm}
\author{Marco Ancona}
\affiliation{SUPA, School of Physics and Astronomy, University of Edinburgh, Peter Guthrie Tait Road, Edinburgh, EH9 3FD, UK}
\author{Alessandro Bentivoglio}
\affiliation{SUPA, School of Physics and Astronomy, University of Edinburgh, Peter Guthrie Tait Road, Edinburgh, EH9 3FD, UK}
\author{Michele Caraglio}
\affiliation{Institut für Theoretische Physik, Universität Innsbruck, Technikerstraße 21A, A-6020 Innsbruck, Austria}
\author{Giuseppe Gonnella}
\affiliation{Dipartimento di Fisica, Universit\`a degli Studi di Bari and INFN, Sezione di Bari, 70126 Bari, Italy}
\author{Alessandro Pelizzola}
\affiliation{Dipartimento Scienza Applicata e Tecnologia, Politecnico di Torino, Corso Duca degli Abruzzi 24, 10129 Torino, Italy}
\affiliation{INFN, Sezione di Torino, via Pietro Giuria 1, 10125 Torino, Italy}

\date{\today}

\begin{abstract}
\vspace{-0.3 cm}

We investigate the possibility of extending the notion of temperature in a stochastic model for the RNA/protein folding driven out of equilibrium. We simulate the dynamics of a small RNA hairpin subject to an external pulling force, which is time-dependent. First, we consider a fluctuation-dissipation relation (FDR) whereby we verify that various effective temperatures can be obtained for different observables, only when the slowest intrinsic relaxation timescale of the system regulates the dynamics of the system. Then, we introduce a different nonequilibrium temperature, which is defined from the rate of heat exchanged with a weakly-interacting thermal bath. Notably, this `kinetic' temperature can be defined for any frequency of the external switching force. We also discuss and compare the behavior of these two emerging parameters, by discriminating the time-delayed nature of the FDR temperature from the instantaneous character of the kinetic temperature. The validity of our numerics are corroborated by a simple 4-state Markov model which describes the long-time behaviour of the RNA molecule.

\end{abstract}

\maketitle

\section{I. Introduction}

Many natural and physical systems evolve under nonequilibrium conditions.
They can be living or biological systems where chemical energy is continuously converted in movement or mechanical work, or slow processes where relaxation times to equilibrium exceed measurable timescales. 
Statistical physics, from its foundation, has always tried to conceive a theoretical framework for the study of nonequilibrium systems.
Yet, a list of general results akin to those existing for the equilibrium counterparts is still lacking.
Recently, fluctuation relations~\cite{Gallavotti1995JSM,Lebowitz1999,seifert2005,Ritort2008,Bustamante2005,jarzynski2011,Seifert2012,gradenigo2013}  and macroscopic fluctuation theories~\cite{Bertini2001,Bertini2015} have provided major advances in the statistical description of nonequilibrium phenomena.
However, a substantial gap between our current understanding of nonequilibrium fundamentals and what we know for equilibrium still remains.

One of the most established concepts in equilibrium thermodynamics and statistical mechanics is temperature. 
Temperature has a genuine statistical origin, as it represents the average kinetic energy in large systems with several degrees of freedom. 
When in contact with a second system (often a thermal bath), temperature regulates heat exchanges between the two. 
Extending this notion to the nonequilibrium context is one of the grand challenge of the current theoretical approaches to nonequilibrium physics.
For glassy systems, which display nonequilibrium aging properties, mean-field models and simulations suggest the emergence of an equilibrium-like temperature, defined via a relation similar to the fluctuation-dissipation theorem (FDT)~\cite{Cugliandolo2000,Speck2006,seifert2010,Cugliandolo2011}.
The idea is to identify the parameter that replaces the bath temperature in a fluctuation-dissipation relation (FDR)~\cite{gradenigo2013} between the time-delayed correlation and the linear response of the same observables as an effective temperature.

More precisely, one exploits the relation (setting $k_B = 1$):
\begin{equation}
T_{eff}(\Delta t)\chi_\mathcal{O}(\Delta t)= C_\mathcal{O}(\Delta t) \; ,
\label{FDTintro}
\end{equation}
where the \textit{self-correlation} $C_{\mathcal{O}}$ quantifies the spontaneous fluctuations of a given observable $\mathcal{O}$ and $\chi_{\mathcal{O}}$ is the \textit{integrated linear response function} representing the response of a system to an external perturbation. 
In the long time-delay limit, $\Delta t >> t_c$, being $t_c$ some transient timescale, many interesting systems, including those with aging dynamics~\cite{Cugliandolo1997,cugliandolo1997energy,Puglisi2017,nandi2018,Montanari2003}, active matter~\cite{loi2008,Levis2015,palacci2010,gradenigo2013,suma2014,szamel2014,patteson2016,preisler2016,szamel2017,Petrelli2018,Petrelli,Petrelli2020,Flenner2020}   and polymer physics~\cite{loi2011}, reach a regime in which $T_{eff}(\Delta t)$ saturates to a constant $T_{eff}$ that under certain conditions
can be interpreted as an effective temperature regulating all thermal and heat exchange properties of the system~\cite{cugliandolo1997energy,Cugliandolo2011,Puglisi2017,loi2011}.

Despite this, the possibility of defining an effective temperature for many classes of non-equilibrium  systems is still to be assessed. 
Only few experiments support the validity of the effective temperature notion, while many theoretical and numerical results raise important questions on the real meaning of such quantity, by inspecting its dependence on the specific considered observable~\cite{Droz2009,Levis2015}, or asking whether it has a relevant role in regulating the nonequilibrium thermodynamics~\cite{Baiesi2009}. 
Therefore, it could be useful to reconsider the concept of effective temperature in some simple but realistic model where timescales are under control.

Small fluctuating systems offer a convenient possibility to investigate on the role of effective temperature, since they are completely characterized in equilibrium conditions, and their study is still feasible when driven out of equilibrium~\cite{Seifert2012}. 
An important example of such category is represented by small biopolymers, such as RNA or DNA fragments, and short proteins.
They can adopt different structural conformation under some environmental conditions (bath temperature, salt concentrations, external pulling forces etc.).
Such small molecules can be often equivalent to a two-state system, as they can be in either a folded configuration or an unfolded conformation.
In such systems, a possible pathway towards nonequilibrium is to force the folding-unfolding transitions by an external random force, which prevents the system to equilibrate.
In particular, one can ask how the folding-unfolding dynamics of proteins/RNA molecules are affected by this external drive, and whether the nonequilibrium properties can be characterized by the effective temperature mentioned above.
Recently, the emergence of an effective temperature in randomly pulled biomolecules has been experimentally ascertained by Dietrich \textit{et al}.~\cite{Ritort2015}.
By going in this direction, an analysis of the typical relaxation timescales and a comparison of the fluctuations (correlations) of the various observables in such class of systems can help to shed light on the role of the effective temperature.

In this paper, we consider a model, originally introduced in~\cite{WS1,WS2,ME1,ME2,ME3}, that can realistically reproduce equilibrium and dynamic behaviors of small RNA molecules and proteins.
In the context of equilibrium, this model has been exactly solved in references~\cite{Pelizzola2002,Pelizzola2005,pelizzola2005exact}, and successfully used to predict the equilibrium and dynamical behavior of several biomolecules~\cite{ZP2006-1,ZP2006-2,BPZ2007-1,BPZ2007-2,IPZ2007,imparato2007,imparato2008,Imparato2009,ZP2009,CIP2010,caraglio2011,FBP2011,caraglio2012,PZ2013,FBP2015}. 
Here, we use this model to examine the nonequilibrium properties of an RNA hairpin: we measure integrated correlation and response functions of different observables, and we evaluate the typical relaxation timescales which play an essential role in determining the emergence of an effective temperature.
Our results are broadly in line with the experimental findings in~\cite{Ritort2015}.
Then, we also compare the effective temperature defined via the FDT-like relation in Eq.~\eqref{FDTintro} with another `kinetic' temperature, which quantifies the extent of heat exchanged between the RNA fragment and a weakly--coupled system at a different temperature. 

The paper is organized as follows.
In Section II we define the model used, and we briefly describe the main feature of the RNA fragment that we have analyzed.
An outline of the main results on the equilibrium properties of this molecule are shown in subsection IIA.
In subsection IIB we present preliminary simulations in nonequilibrium conditions. We show some representative time series of the system, describing its qualitative response to the external random force.
In Section III, we recall the rudiments of the fluctuation-dissipation relation (FDR) out of equilibrium, and we introduce two possible nonequilibrium temperatures for our system. In section IV, we develop an analytically solvable 4-state model, which poses the guidelines to understand our numerics. Then, the simulation results on the effective temperature calculated via the FDR are presented in Section V, for a large range of the relevant parameters; there, we compute the effective temperature for two different variables, the end-to-end length of the molecule, $L$ (subsection VA), and the number of native contacts $N_c$ (subsection VB). A detailed discussion on the relevant timescales of this system is proposed throughout the whole section, by means of a direct comparison with the 4-state model predictions. In the subsection VC, we evaluate a kinetic temperature for our model. Therein, we also discuss the analogies and differences with the FDR effective temperature.

\section{II. Model and Methods}

A $N$-residues-long protein/RNA is modeled as a 1D lattice of $N+2$ sites, where the $N$ bulk sites represent the residues/bases and the boundary sites are the terminal ends.
Each site is labelled by a dichotomous variable $m_k$, with $k = 1, \ldots, N$, which describes its nativeness: if $m_k = 1$ the $k$-th residue is native, while if $m_k = 0$ it is not.
Boundary conditions are specified by $m_0 = m_{N+1} = 0$. 
Similarly, any segment of the molecule enclosed within the $i$-th and $j$-th site can be native or nonnative. 
A native $ij$-stretch is defined as a sequence of consecutive native residues ($m_k = 1$ for $k = i+1,j-1$) delimited by two nonnative sites at the boundaries ($m_i = m_j = 0$). 
Then, the auxiliary variable $S_{ij} \equiv (1-m_i)(1-m_j)\prod_{i+1}^{j-1} m_k$ is linked to the nativeness of stretches, being equal to $1$ if the $ij$-stretch is native and $0$ otherwise.
Due to the 3D folding of the protein/RNA chain, in a folded structure, each atom of a residue $i$ is in contact with the atoms of another residue $j$ if their distance is lower than a threshold distance that we set equal to 4\AA.
The number of atomic contacts and the distances in three-dimensional real space between residues in the folded structure are given respectively by the matrix elements $n_{ij}$ and $l_{ij}$. 
Such matrices are input values of the model, depend on the particular protein/RNA considered, and are extracted from the relative file in the Protein Data Bank (PDB)~\cite{PDB}.
We assume that each atomic contact is associated with an energy term $-\epsilon$, so that a pair of residues $(i,j)$ with $n_{ij}$ atomic contacts will contribute to the total energy with an energetic loss of $-\epsilon n_{ij}$, when the molecule is in its native configuration. 
In the same condition, if an external constant force $f$ acts on one terminal end of the chain, a further energetic contribution comes from the term $ -f l_{ij} \sigma_{ij}$, where $\sigma_{ij} = \pm 1$ is another binary variable of the model representing the orientation of the $ij$-stretch with respect to the force direction. 
Given a particular configuration $(\{m_k\}$,$\{\sigma_{ij}\})$, we define
\begin{equation}
N_c \equiv \sum_{i=1}^{N-1} \sum_{j=i+1}^{N} n_{ij}\prod_{k=i}^{j} m_k \; ,
\label{Nc}
\end{equation}
which represents the total number of native atomic contacts, while 
\begin{equation}
L \equiv \sum_{i=0}^{N+1} \sum_{j=i+1}^{N} l_{ij} S_{ij} \sigma_{ij} \; 
\label{L}
\end{equation}
is the end-to-end length.

In the presence of a constant pulling force $f>0$, the equilibrium properties of the RNA/protein can be described by its Hamiltonian: 
\begin{equation}
\mathcal{H} = -\epsilon N_c - f L \; .
\label{hamiltonian}
\end{equation}
We assume that only nativelike residues which belong to the same native stretch can lower the energy of the system. This is encoded in the product $\prod_{k=i}^{j} m_k $ in Eq.~\eqref{Nc}, which is nonzero only if $m_k = 1$ holds for $k = i,i+1,..,j-1,j$. In such way, we mimic the cooperative folding in real protein/RNA molecules.
Similarly, we assume that only native stretches ($S_{ij} = 1$) contribute to the end-to-end length $L$, as can be seen in Eq.~\eqref{L}. 
For instance, if the molecule is kept at zero temperature and small force, the equilibrium configuration is the one with all the bulk residues native ($m_i = 1$ for every $i = 1,\ldots,N$ and $S_{0,N+1} = 1$), which means that the whole molecule is in the native conformation, and its effective length is the folded length measured by experiments.
For a system in contact with a thermal bath at a finite temperature $T$, each configuration $(\{m_k\}$,$\{\sigma_{ij}\})$ can be visited by the system, with a probability which is only proportional to the Boltzmann weight $\text{exp}(-\beta \mathcal{H})$, where $\beta = 1/T$ ($k_B = 1$).
Therefore, $T/\epsilon$ and $f/\epsilon$ are the control parameter at equilibrium, while out of equilibrium $\epsilon$, $f$ and $T$ will be considered separately (see below).

\begin{figure}[t]
\includegraphics[width=0.5\textwidth]{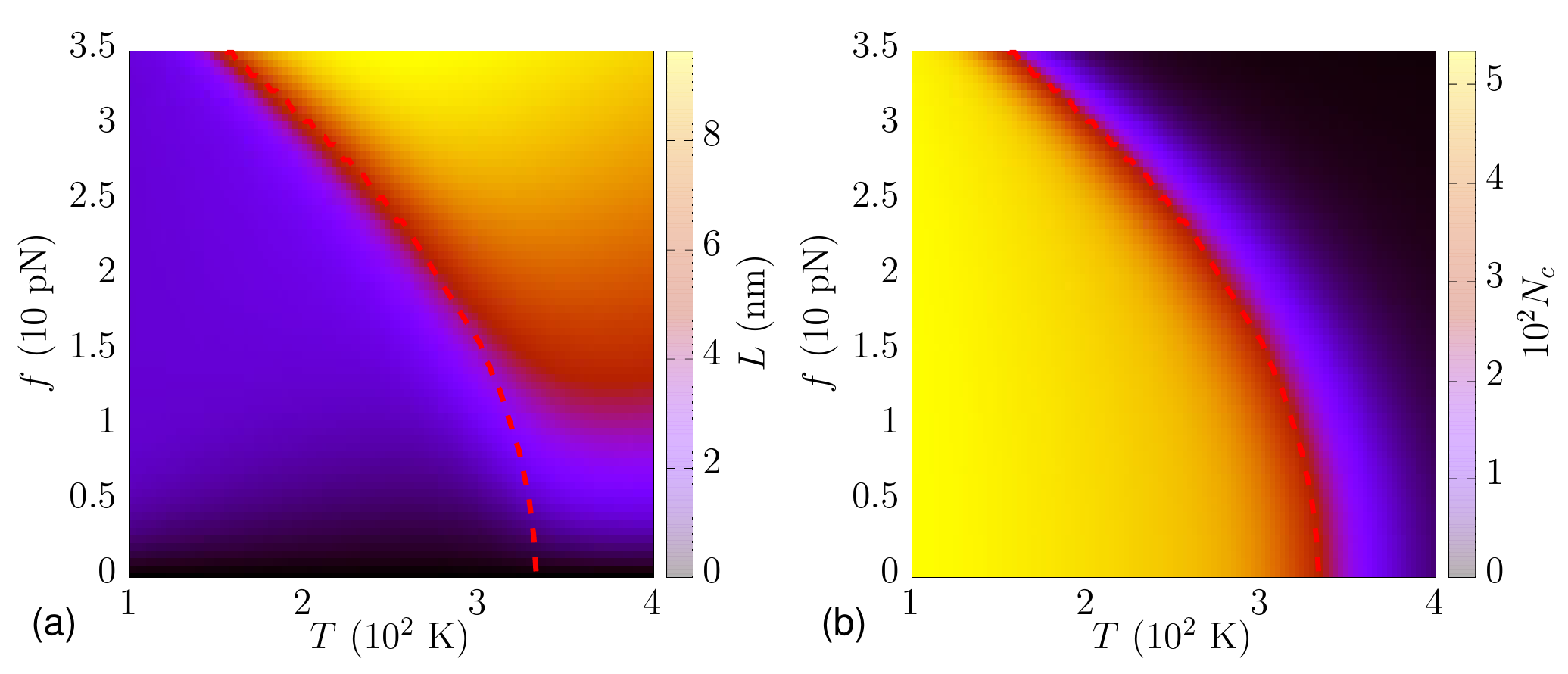}
\caption{\textbf{Equilibrium phase diagrams for $\mathbf{L}$ and $\mathbf{N_c}$.} In this figure we show the $L$ (panel (a)) and $N_c$ (panel (b)) values in the $f$--$T$ space (fixed $\epsilon$). We also draw the crossover line (bright red dashed line), which consists of the points in the $f$--$T$ diagram for which $2/3$ of the nucleotides are nativelike. This is the criterion used in~\cite{Imparato2009} to individuate the folding-unfolding crossover at different bath temperatures. \textbf{(a)} This panel shows the crossover between two different regimes. In the yellow region, the molecule is fully extended, and oriented towards the force direction. In the purple/black region the RNA chain is either in the hairpin configuration, and thus it is folded (below the crossover line), or it is unraveled, but does not align with the force (bottom--right corner of this panel). For both cases, $L$ is below the value $L \sim 5$nm. \textbf{(b)} The number of native contacts $N_c$ correctly predicts the order-disorder transition for this RNA. Indeed, the crossover line between the native/ordered configuration and the nonnative/disordered one locates in the red region, which corresponds to $1/2$ of the native contacts to be nativelike ($N_c \simeq 250$).}
\label{fig1}
\end{figure}

\begin{figure*}[t]
\includegraphics[width=1.\textwidth]{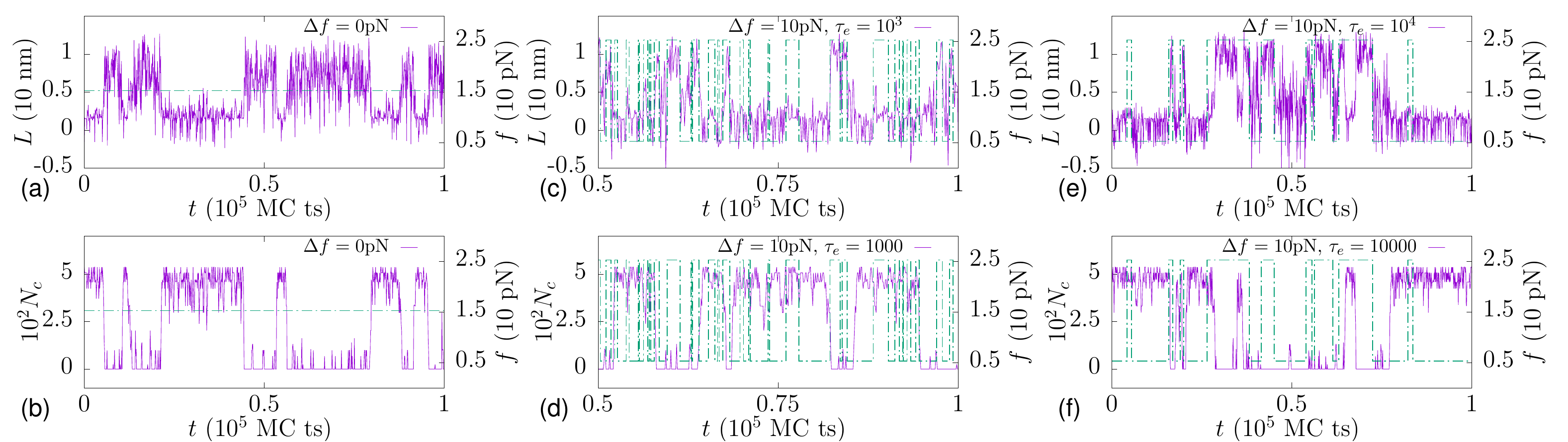}
\caption{\textbf{Time series of in and out of equilibrium RNA.} Representative time series of the relevant observables $L$ and $N_c$ (purple lines) and typical force profiles (green lines), for equilibrium and $\tau_e = 1000,10000$, see Eq.~\eqref{telegraph}. The simulations were run for $T = 300$, $f_{bias} = f_c = 15.3$pN, $\Delta f = 0$ (panels (a,b)) and $\Delta f = 10$pN (panels (c,d,e,f)), $\epsilon = 13.92$. \textbf{(a,b)} At equilibrium, $L$ and $N_c$ switch with a typical timescale of the system at this temperature. Large fluctuations in $L$ are manifest in the unfolded basin. \textbf{(c,d)} For small $\tau_e$, the switching force produces a decrease in the typical transition time of $L$ and $N_c$ between the two states. However, the dynamics of observables do not follow force jumps. \textbf{(e,f)} Conversely, for large $\tau_e$, the RNA molecule is able to respond to the intermittent hops of the stochastic force. Thus, $L$ trajectory tend to mimic the force profile, while $N_c$ trajectory anticorrelates with force values.}
\label{fig2}
\end{figure*}
In this work we have simulated the 22-nucleotides PG5A RNA hairpin, for which the input data needed, $n_{ij}$ and $l_{ij}$, can be extracted from the PDB file in \cite{PDB} (code 1F9L).
The dynamical properties of this and similar RNA hairpins have been widely studied both experimentally~\cite{Liphardt2001} and numerically~\cite{Hyeon,Kyu,Imparato2009}, at equilibrium and under nonequilibrium conditions.
Here, instead, we focus on the thermodynamic properties of this RNA segment, mainly to illustrate the emergence of effective temperatures in nonequilibrium conditions. 
This model has been used to successfully describe in and out-of-equilibrium dynamical properties 
of several other proteins (such as protein PIN1~\cite{imparato2007}, ubiquitin~\cite{imparato2008}, fibronectin~\cite{CIP2010}, and GFP~\cite{caraglio2011}), and can be used to further explore the emergence of effective temperature in more complicated proteins/RNA chains.
However, in this paper we restrict ourselves to the analysis of the PG5A RNA hairpin, as it is instructive to comprehensively illustrate the emergence of nonequilibrium temperatures in wide ranges of parameters, which may not be feasible for systems with a larger number of degrees of freedom.

To investigate the stochastic dynamics of the PG5A RNA hairpin, we perform Monte Carlo simulations. We consider a time-dependent random force $f(t)$ that switches intermittently between the two values $f_{bias} \pm \Delta f$, with a typical switching time $\tau_e$, or, equivalently, such that 
\begin{equation}
\begin{split}
\langle f(t) \rangle &= f_{bias} \; , \\
\langle f(t)f(t') \rangle = f_{bias}^2 &+ (\Delta f)^2 e^{-2|t-t'|/\tau_e} \; ,
\end{split}
\label{telegraph}
\end{equation}
that are respectively the mean value and the covariance of a two-state telegraph process for symmetric jumps about the bias~\cite{Gardiner}. In the algorithm, at each Monte Carlo time step (MC ts), the force value can switch with a rate $1/\tau_e$. The equilibrium condition is met when $\Delta f = 0$.

In simulation in and out of equilibrium, the state of both a randomly chosen $k$-th site and an $ij$-stretch can modify as follows: $m_k \to 1-m_k$, $\sigma_{ij} \to -\sigma_{ij}$, according to the Metropolis rule.
The simulations were equilibrated for $5 \cdot 10^4$ time step, and then were run for at least $2 \cdot 10^4$ time steps.

\subsection{A. Equilibrium properties of PG5A RNA hairpin}
In equilibrium conditions (i.e. $f$ constant), the system displays a folding-unfolding crossover~\cite{Hyeon,Imparato2009}.
In terms of the nativeness of the nucleotides, this crossover can be characterized by the mean number of nativelike residues $m \equiv (1/N)\sum_{i=1}^{N} \langle m_i \rangle$. 
When RNA is stable in the native configuration (small $T,f$), the order parameter $m$ is approximately $1$, whilst in the totally disordered RNA ($T$ large) $m$ is about $1/3$~\cite{Imparato2009}.
Therefore, the folding-unfolding crossover line can be individuated for those force and temperature values for which $2/3$ of the residues are native.
Such criterion has been used to locate in the $f$--$T$ diagram the crossover points between the ordered and the disordered macrostates, and to find the correspondent energy landscapes~\cite{Imparato2009}.
In Fig.~\ref{fig1}, we report the crossover line found by following this criterion.
We observe that the end-to-end length $L$ cannot be used  to individuate the crossover between the ordered/folded regime and the disordered/unfolded one. Indeed, for high temperatures and low forces, such observable is not able to capture the nativeness of the RNA structure.
Indeed, for large $T$ and small values of $f$, the probability distribution associated with $L$ is symmetric and centered in $L \simeq 0$ (not shown), yielding a mean value similar to the one in the ordered phase.
This is shown in Fig.~\ref{fig1}(a), where $L$ values are plotted in the $f$--$T$ space.
Clearly, for $T \gtrsim 333 K$, $f \lesssim 8$pN, there is a deviation of the red region, which signals intermediate values of $L$, from the real crossover line obtained with the aforementioned criterion.
Conversely, a good order parameter which describes this crossover is the number of native contacts $N_c$.
In Fig.~\ref{fig1}(b) we show the total number of native atomic contacts in the $f$--$T$ space.
Note that the phase diagram is qualitatively similar to the one shown in~\cite{Imparato2009} for $m$ (compare to Fig.~1 in that paper), with a sharp crossover between the native/folded and the nonnative/unfolded macrostates of the RNA molecule.
Moreover, the crossover line overlaps with the red region in the phase diagram ($1/2$ of contacts are native).
Both diagrams in Fig.~\ref{fig1} are obtained by analytical calculations, since partition function, and thus mean values of any quantities, can be exactly computed by means of Eq.~\eqref{hamiltonian}, as demonstrated in Ref.~\cite{Pelizzola2002}. The value of $\epsilon$ is equal to $13.92$, that is the temperature scale factor which reproduces the experimental critical unfolding temperature in the absence of a pulling force ($T_c = 333K$). 

We finally remark that, due to the finite length of the PG5A RNA chain, such folding-unfolding transition shows up as a sharp crossover between two macrostates, with a marked bistability in proximity of the crossover line.
Indeed, thermal-induced transitions between the folded/ordered and the unfolded/disordered phases occur at the unfolding force $f_c = 15.3$pN, as shown in the representative time series in Figs.~\ref{fig2}(a,b), respectively for $L$ and $N_c$. This corresponds to the crossover value reported in Refs.~\cite{Hyeon,Imparato2009} and Fig.~\ref{fig1} at the bath temperature $T = 300 K$.
The crossover line in Fig.~\ref{fig1} is interpreted as a real order-disorder transition line in the thermodynamic limit, where the order parameter $m$ (or $N_c$) can exhibit a discontinuous jump at the transition values of the control parameters $f$ and $T$. 
Therefore, in the rest of the paper we will refer to the crossover line and the unfolding force $f_c$, by unambiguously using terms as `transition line' and `critical force'. 

From the timeseries in Figs.~\ref{fig2}(a,b), it is also possible to find a rough estimation of the conversion factor between Monte Carlo and real time units, at equilibrium. Comparing the real unfolding/refolding times of the PG5A RNA hairpin given in \cite{Hyeon} to the residence times calculated in our model, we find that $1$ Monte Carlo time step corresponds to about $10^{-4}$--$10^{-3}$ ms.

\subsection{B. Out-of-equilibrium dynamics of PG5A RNA hairpin}

We now switch to a nonequilibrium context,
i.e. $f$ is time--dependent as detailed in Section~II, with expectation value and correlation defined as in Eq.~\eqref{telegraph}. 
In Fig.~\ref{fig2} we show the time series of $L$ and $N_c$ in nonequilibrium conditions ($\Delta f = 10$pN), at $T=300 K$, for the representative values of the force timescale, $\tau_e = 1000,10000$.
In the former case, the typical residence times spent in the folded and the unfolded states reduces for both $L$ and $N_c$.
In such conditions, those are also associated with the `longest' relaxation timescale of the system, or, in other words, the time that the system needs to uncorrelate from its initial state.
However, since the molecule is not able to respond immediately, for such value of $\tau_e$ the RNA dynamics differs significantly from the force time profile (see Figs.~\ref{fig2}(c,d)). 
Conversely, for $\tau_e = 10000$, the switching dynamics follows the force dynamics, since the system has enough time to respond to the force jumps. In Figs.~\ref{fig2}(e,f) is clearly shown that the end-to-end length (number of native contacts) time series is correlated (anticorrelated) with the force time profile.
For large $\tau_e$, the `longest' relaxation time is approximately $\tau_e/2$, as we will show below. 
\break
\section{III. Nonequilibrium temperatures}

The characterization of the thermodynamic state of an out-of-equilibrium system via an effective temperature is an attempt to understand a nonequilibrium problem into an equilibrium framework. 
In equilibrium conditions, all the definitions of $T$ lead to the same outcome, which is usually the bath temperature, as this measure is uniquely related to the mechanism of heat dissipation, which is the only factor that governs the dynamics. Generally, this latter consideration does not hold out of equilibrium, and, thus, a comparison between different temperature definitions is in order.

In this section we define two different effective temperatures which will be calculated for our model, the FDR temperature, $T_{FDR}$, and the \textit{kinetic temperature}, $T_{kin}$ respectively. The two definitions inform about two different aspects of nonequilibrium systems: while the former is more related to the time-delayed properties of the systems (which are quantified by two-times correlation and response functions), the latter is rather associated with the instantaneous exchange of heat in the nonequilibrium stationary state.

\subsection{A. FDR effective temperature}

In order to introduce the FDR for our model, we need to define the \textit{integrated correlation} function and the \textit{integrated linear response} function.
Suppose that $X$ is a generic observable of the system, which assumes the value $x(t)$ at time $t$, and the system is described by the Hamiltonian $\mathcal{H}_0 - g(t)X$, where $g(t)$ is the time-dependent intensive variable conjugated to $X$. At time $t_0 = 0$ a small steplike perturbation $\delta g$ is applied, such that $\mathcal{H}_{t>t_0} = \mathcal{H}_0 - [g(t)+\delta g]X$.
Thus, the integrated correlation and response functions are given by:
\begin{equation}
C_X(t) \equiv \langle \left[x(t_0) - x(t)\right]x(t_0) \rangle_{ss} \; ,
\label{corr}
\end{equation}
\begin{equation}
\chi_X(t) \equiv \frac{\langle x(t) - x(t_0) \rangle}{\delta g}, \qquad t \geq t_0 \; ,
\label{resp}
\end{equation}
where the symbol $\langle ..\rangle_{ss}$ denotes the expectation value in the nonequilibrium steady state (NESS), while $\langle .. \rangle$ is the expectation value computed in the presence of the small perturbation $\delta g \to 0$. Note that $C_X$ and $\chi_X$ are monotonically increasing functions of time, which satisfy $C_X(t_0) = \chi_X(t_0) = 0$, and $\chi_X(\infty) \equiv \chi_\infty$, where $\chi_\infty$ is the (asymptotic) susceptibility. At equilibrium, they are strictly related by the FDT, which in its integrated version reads as follows:
\begin{equation}
\frac{\chi_X(t)}{C^{eq}_X(t)/T} = 1, 
\label{FDT}
\end{equation}
where $T$ is the bath temperature. The superscript `$eq$' means that the average has to be performed in the equilibrium steady state.
Moreover, at equilibrium, Eq.~\eqref{FDT} works for any variable $X$ at any time $t > t_0$.
Such theorem is violated out of equilibrium.
In spite of this, a relation similar to Eq.~\eqref{FDT} can be written also for systems in their nonequilibrium steady state: 
\begin{equation}
Y(t) \equiv  \frac{\partial \chi_X(t)}{\partial (C_X(t)/T)}.
\label{FDR}
\end{equation}
Eq.~\eqref{FDR} represent a formulation of FDR, where $Y(t)$ is the \textit{violation parameter}~\cite{Cugliandolo1997,Baiesi2009}.
$Y(t)$ is the slope of the parametric curve $\chi_X(C_X/T)$ in Eq.~\eqref{FDR} at each time $t > t_0$; thus, in general, the aforementioned parametric function displays a nonzero curvature. 
Nonetheless, for a large class of systems and observables, such factor is independent of $t$ after some time threshold $\tau_c$ (see Section IV), and an \textit{effective temperature} $T_{FDR}$ can be defined, such that 
\begin{equation}
Y \equiv T/T_{FDR}, \qquad t \gg \tau_c \; .
\label{efftemp}
\end{equation}
In this latter case, and FDT-like relation is restored by substituting in Eq.~\eqref{FDT} the bath temperature $T$ with the parameter $T_{FDR}$. Clearly, the equilibrium limit verifies $Y(t) = 1$ and $\tau_c = 0$, which implies Eq.~\eqref{FDT}.

Eqs.~\eqref{corr} and \eqref{resp} can be calculated either in equilibrium or in nonequilibrium conditions, as long as an unique steady state exists.
The integrated response function can be computed much more easily than the usual response function in numerical simulations.
Therefore, from now on, we will only use $C_X(t)$ and $\chi_X(t)$ as a measure of the correlations and the response to a perturbation for the observables $L$ and $N_c$ introduced in Section II.

\subsection{B. The kinetic temperature}

In this subsection, we introduce a nonequilibrium temperature of a different nature, which we will refer to as `kinetic' temperature, $T_{kin}$. Its definition is based on the rate of heat that would be exchanged between the whole system and another (virtual) thermal bath, which serves as a `thermometer'. Then, we explain a simple and computationally efficient way to evaluate such kinetic
temperature.

We imagine our system to be in contact with a second weakly interacting bath, at temperature $T_{th} \neq T$. Therefore, this second bath is virtually able to exchange heat with the system (or equivalently with a subset of degrees of freedom) without modifying its state. The rate of heat that would be absorbed is, on average:
\begin{equation}
\langle \dot Q_X \rangle = \sum_{\Sigma} \sum_{\Sigma' \in \partial_X \Sigma} P^0(\Sigma) [E(\Sigma') - E(\Sigma)] W_{\Sigma,\Sigma'} \; . 
\label{kin}
\end{equation}
$X$ is the variable (or the set of variables) weakly coupled with the second thermal bath at temperature $T_{th}$ and $\Sigma$ indicates the global state of the system (in our case, it is determined by the microscopic variables $\{ m_k, \sigma_{ij}\}$).
$P^0(\Sigma)$ is the NESS probability distribution associated with $\Sigma$. $E(\Sigma)$ is the energy of the system when in the state $\Sigma$, and $W_{\Sigma,\Sigma'}$ is the transition rate from the state $\Sigma$ to the state $\Sigma'$. $\partial_X \Sigma$ is the set of states which can be reached in those transitions that modify only the variable $X$. For models where it is not possible to split all possible transitions into subsets regarding different observables, one can still define a single kinetic temperature by considering all possible transitions in the second summation in Eq.~\eqref{kin}. The dependence on $T_{th}$ is implicit in the transition rates $W_{\Sigma,\Sigma'}$, whilst $P^0(\Sigma)$ and $E(\Sigma)$ are independent of $T_{th}$, since the second bath is only weakly--interacting. For the Metropolis dynamics, we have $W_{\Sigma, \Sigma'} \equiv \text{min}\{ 1, e^{-\beta_{th} \left[E(\Sigma') - E(\Sigma)\right]} \}$, where $\beta_{th} = 1/T_{th}$. Observe that the heat exchanged per unit of time, $\langle \dot{Q}_X \rangle$ can depend on the particular observable $X$. Reasonably, the thermometer measures the effective temperature $T_{kin}$ of the system when $\langle \dot{Q}_X \rangle = 0$, or, in other words, the second thermal bath will be at temperature $T_{th} = T_{kin}$ when no heat is exchanged (on average). This latter condition defines the kinetic temperature of the system.

Thus, the kinetic temperature can be operatively computed in a simulation run in its NESS, as follows: 
\begin{itemize}
\item[(i)] The energy $E(\Sigma)$ is calculated at each time step;
\item[(ii)] If a transition occuring at time $t$, $\Sigma \rightarrow \Sigma'$, modifies the value (or values) of the variable (or the set of variables) $X$, the variation in energy $E(\Sigma') - E(\Sigma)$ is stored;
\item[(iii)] then, for some temperature $T_{th}$, every variation in energy of the type in (ii) is weighted with the corresponding transition rate $W_{\Sigma,\Sigma'}$ and the summation in Eq.~\eqref{kin} is performed;
\item[(iv)] the temperature $T_{th}$ is systematically varied, and the procedure in (iii) repeated to calculate $\langle \dot{Q}_X(T_{th}) \rangle$, until the condition $\langle \dot{Q}_X(T^*_{th}) \rangle \approx 0$ is met. The value $T^*_{th}$ estimates $T_{kin}$.
\end{itemize}

Note that for such procedure to be applicable, the prior knowledge of the transition rates $W_{\Sigma,\Sigma'}$ is needed. For a Monte Carlo dynamics, for example, such requirements are always satisfied. In subsection VC, we calculate the kinetic temperatures relative to the microscopic observables $m_i$ and $\sigma_{ij}$.

\section{IV. 4-STATE MODEL}

To better understand the numerical results presented in the following section, we map our RNA into a simpler system, which can be either in the folded/ordered state or in the unfolded/disordered one, following the effective 4-state model described in \cite{Ritort2015}. We remark that this framework is generic and indeed our analytical predictions hold for any 4-state system that follows the same transition rules.

The observable that describes the system is labelled by $s=s_\pm$, and it is forced by an external two-state drive, labelled by $x=x_\pm$. 
The states of this effective 4-state model are $(s,x) \equiv \{1,2,3,4\} = \{(s_+,x_+),(s_+,x_-),(s_-,x_+),(s_-,x_-) \}$ and the master equation which governs the dynamics is:
\begin{equation}
\partial_t \mathbf{P}(t) = \mathbb{M} \mathbf{P}(t) \; ,
\end{equation}
where $\mathbf{P}(t) \equiv \mathbf{P}_{(s,x)}(t)$ is a 4-state probability vector, such that $\sum_{s = s_\pm ,x = x_\pm} P_{(s,x)}(t) = 1 $, at every time $t$. 
The matrix element $M_{ij}$ is the transition rate from state $i$ to state $j$. 
Thus, the matrix $\mathbb{M}$ reads:
\begin{equation}
\mathbb{M} = 
\begin{pmatrix}
M_{11} & 1/\tau_e & M_{13} & 0 \\
1/\tau_e & M_{22} & 0 & M_{24} \\
M_{31} & 0  & M_{33} & 1/\tau_e  \\
0 & M_{42} & 1/\tau_e & M_{44}.
\end{pmatrix}
\label{matrix}
\end{equation}
given that $M_{jj} = - \sum_{i, i\neq j} M_{ij}$, with $i,j = 1,2,3,4$. Eigenvalues $\lambda_k$ and right (left) eigenvectors $\mathbf{P}^k$ ($\mathbf{Q}^k$) of $\mathbb{M}$ are such that $ \mathbb{M}\mathbf{P}^k = \lambda_k \mathbf{P}^k$  ($\mathbf{Q}^k\mathbb{M} = \mathbf{Q}^k \lambda_k $), for $k=0,1,2,3$. The quantities $\mu_k = -\lambda_k$ are nonnegative and represent the inverse of the typical timescales of the system. Since the system reaches the steady state eventually, we have that $\mu_0 = 0$, and the corresponding right eigenvector is $\mathbf{P}^0$, the stationary probability distribution. Then, for every $k > 0$, $\tau_k \equiv 1/\mu_k$ defines the timescales of the system. One finds:
\begin{equation}
\begin{aligned}
\begin{gathered}
\mu_1 = \frac{2}{\tau_e} \\
\mu_{(2,3)} = \left(\frac{1}{\tau_e} + \frac{M_{12}+M_{21}+M_{34}+M_{43}}{2} \right) \\
	  \pm \left[\frac{1}{\tau^2_e} + \frac{(M_{34}-M_{12}+M_{34}-M_{21})^2}{4} \right]^{\frac{1}{2}}.
\label{eigenvalues}
\end{gathered}
\end{aligned}
\end{equation}
Correlation and response function are defined as in Eqs.~\eqref{corr} and \eqref{resp}:
\begin{gather}
   C(t) = \langle s_0s_0 \rangle - \langle s_0s_t \rangle, \\
\chi(t) = \frac{\partial \langle s_t - s_0 \rangle}{\partial (\delta g)}\Bigg|_{ \delta g = 0}
\end{gather}
being $s_t$ the value assumed by the stochastic variable in exam at time $t$, and $\delta g$ the small step-like perturbation applied to the bias of the external drive labelled by $x$. After some calculations, both correlation and response functions can be written as a combinations of the components $P^k_{(s,x)}$ of the eigenvectors of $\mathbb{M}$:
\begin{gather}
   C(t) = \sum_{k=1}^{3} \left(\sum_{s,x} s P^k_{(s,x)} \right) \Gamma_k (1-e^{-\mu_k t}) \label{Cmodes}, \\
\chi(t) = \sum_{k=1}^{3} \left(\sum_{s,x} s P^k_{(s,x)} \right) \gamma_k (1-e^{-\mu_k t}) \label{Rmodes}, \\
\Gamma_k = \sum_{s,x} s Q^k_{(s,x)} P^0_{(s,x)} \label{gammaC}, \\
\gamma_k = \frac{1}{\mu_k} \mathbf{Q}^k \delta \mathbb{M} \mathbf{P^0}. \label{gammachi}
\end{gather}
We now discuss the three timescales $\tau_k$, their relation with the FDR in Eq.~\eqref{FDR} and the existence of an effective temperature. First, note that $\gamma_k$ in Eq.~\eqref{gammachi} depends on $\delta \mathbb{M}$, which represent the first order correction to the transition matrix $\mathbb{M}$ produced by the external perturbation to the NESS.
One can show that $\gamma_1 = 0$ \footnote{Observe that $\mathbf{Q}^1 = (\begin{matrix} 1&-1&1&-1 \end{matrix})$ and, since $\tau_e$ is independent of the perturbation, $\delta M_{21} = \delta M_{12} = \delta M_{43} = \delta M_{34} = 0$. Therefore, for any small perturbation, it follows that $\mathbf{Q}^1 \delta \mathbb{M} = (\begin{matrix} 0&0&0&0 \end{matrix})$, hence $\gamma_1 = 0$}, while $\Gamma_1 \neq 0$. Therefore, on a timescale of the order of $\tau_1 =\tau_e/2$, the ratio $Y(t)/T = \partial \chi(t)/\partial C(t)$ is time-dependent, causing the violation, or curvature, of the FDR in Eq.~\eqref{efftemp} (see also Eqs.~\eqref{Cmodes_approx} and \eqref{Rmodes_approx}). From Eq.~\eqref{eigenvalues}, it is also easy to verify that $\mu_2 > \mu_1$, thus it is always $\tau_2 < \tau_1 $. Therefore, the mode associated with $\mu_2$ in both Eqs.~\eqref{Cmodes} and \eqref{Rmodes} relaxes with a typical time faster than $\tau_1 = \tau_e/2$, which is in turn associated with the curvature term of FDR. This mode converges faster than the violation transient time, and is thus irrelevant for our analysis at large times, see Eq.~\eqref{efftemp}. On a timescale of the order of $\tau_e/2$ or larger, Eqs.~\eqref{Cmodes} and \eqref{Rmodes} reduce to:
\begin{equation}
\begin{split}
   C(t) &\approx A_C (1-e^{-2 t/\tau_e}) +\\
	&+ B_C (1-e^{-t/\tau_3}) \; ,
\end{split}
\label{Cmodes_approx}
\end{equation}
\begin{equation}
\chi(t) \approx A_\chi (1-e^{- t/\tau_3}) \; .
\label{Rmodes_approx}
\end{equation}
where $A_{C,\chi}$ and $B_C$ are prefactors. Thus, for our purposes, two of the three timescales, $\tau_1$ and $\tau_3$, are relevant at large times; in particular, $\tau_3 \equiv \tau_s$ is the slowest intrinsic relaxation timescale of the system under an external perturbation.

From Eqs.~\eqref{Cmodes_approx} and \eqref{Rmodes_approx} one can find the parametric function $C[\chi(t)]$:
\begin{equation}
   C\left[\chi(t)\right] \approx A_C \left[ 1-\left(1-\frac{\chi(t)}{A_{\chi}}\right)^{\frac{2\tau_s}{\tau_e}} \right] + \frac{B_C}{A_{\chi}} \chi(t) \; .
\label{parametricf}
\end{equation}

It emerges that, if $\tau_e$ is sufficiently small, the violation region is restrained to an initial transient, namely the contribution of the first term in the right--hand side of Eq.~\eqref{parametricf} becomes negligible. In particular, this occurs when the curvature $d^2 \chi(C(t))/dC(t)^2$ of Eq.~\eqref{FDR} is about zero. By using Eqs.~\eqref{parametricf} and \eqref{Rmodes_approx}, the inverse curvature can be calculated:
\begin{equation}
   \frac{\partial^2 C}{\partial \chi^2} \approx A_ C\left[\frac{2\tau_s}{A_{\chi}^2 \tau_e} \left(1 - \frac{2 \tau_s}{\tau_e}\right) \mathrm{e}^{-\frac{2}{\tau_s}\left(\frac{\tau_s}{\tau_e} - 1 \right) t} \right]  \; .
\label{curvature}
\end{equation}
Therefore, the condition for a negligible curvature is
\begin{equation}
  t > \frac{\tau_s}{2 \left(\tau_s/\tau_e - 1 \right)} \equiv \tau_c,
\label{tauc}
\end{equation}
which reduces to $\tau_c \approx \tau_e/2$ in the limit of $\tau_e \ll \tau_s$. This is the case only when $\tau_e < \tau_s$, which turns out to be the condition allowing a thermal-like regime at late times (see also Supplementary Material in~\cite{Ritort2015}), as the violation parameter is a constant, $Y(t) \equiv Y$, see Eq.~\eqref{efftemp}.
Vice versa, if $\tau_e > \tau_s$ the curvature in Eq.~\eqref{curvature} is always different from zero, which means that no effective temperature can be detected (or the violation parameter in Eq.~\eqref{FDR} is always time-dependent).

\begin{figure*}[ht]
\includegraphics[width=1.\textwidth]{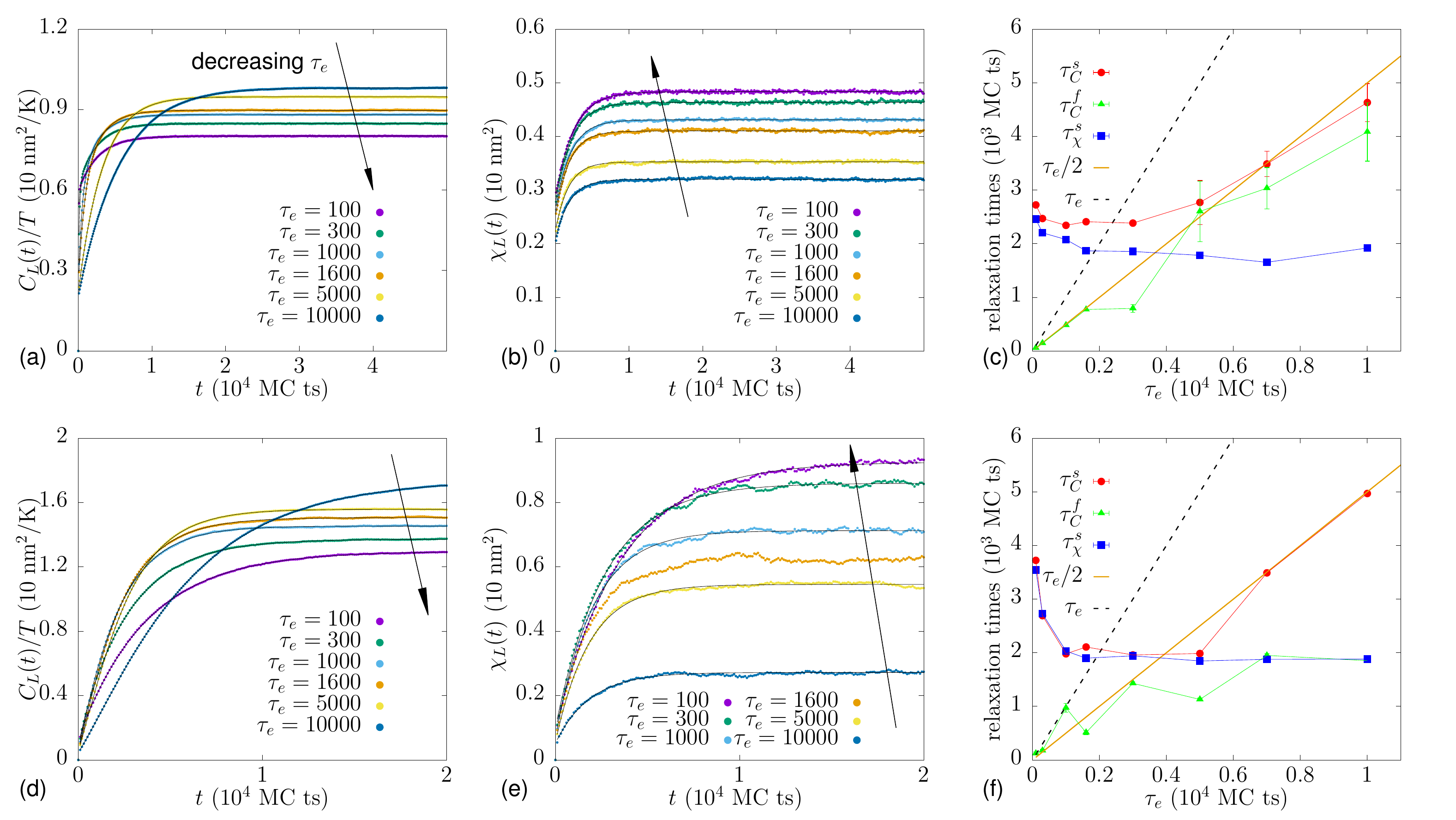}
\caption{\textbf{$\boldsymbol{C_L(t)}$, $\boldsymbol{\chi_L(t)}$ and relaxation timescales.} In this figure, the integrated correlation $C_L(t)$ and response $\chi_L(t)$ curves for different $\tau_e$ are represented for two value of the bath temperature $T=200,300 K$. The response curves are realized by applying a perturbation at $t = 0$. Two values of the perturbation were used to minimize nonlinear effects, $\delta f = 1.0 pN $ ($T = 300K$) and $\delta f = 0.5 pN $ ($T = 200K$) \textbf{(a)} Correlation and \textbf{(b)} response functions for $T=300 K$. The asymptotic values of correlation $C_\infty$ slightly increases as $\tau_e$ increases. Contrarily, the asymptotic susceptibility $\chi_\infty$ decreases with $\tau_e$. \textbf{(c)} Relaxation timescales for $T=300$ are shown as a function of $\tau_e$. The slowest timescale, $\tau^s_\chi$, reaches the plateau at $\tau_e \gtrsim 1000$, which is also the location of the minimum of $\tau^s_C(\tau_e)$. The fast timescale $\tau^f_C$ grows as $\tau_e/2$ for small $\tau_e$. \textbf{(d)} Correlation and \textbf{(e)} response functions for $T=200K$. The integrated response function $\chi$ rapidly decreases with $\tau_e$. \textbf{(f)} Relaxation timescales for $T=200$ are shown as a function of $\tau_e$. The relaxation timescales display the same qualitative behavior as before. For small $\tau_e$, $\tau^s_C$ is more susceptible than in the case $T=300 K$, and $\tau^f_C \simeq \tau_e/2$. For large $\tau_e$, $\tau^s_C$ and $\tau^s_\chi$ approximately match with the ones in panel (c).}
\label{fig3}
\end{figure*}

Such model can be exploited to calculate explicitly the kinetic temperature defined in subsection IVB. Assuming that the external drive which oscillates between the values $x_{\pm} = x_{bias} \pm \Delta x$ is conjugated to the observable labelled by $s_t$, at time $t$, then the energy difference in the right--hand side in Eq.~\eqref{kin} is readily found: $E(s_{t+1},x)-E(s_{t},x) = -x (s_{t+1}-s_{t})$. After some algebra, one finds the following expression of the violation parameter $Y^{kin} \equiv T/T_{kin}$ in the limit of large $\tau_e$:
\begin{equation}
Y^{kin} \simeq 1 - \frac{T}{\Delta x (s_+ - s_-)} \frac{1-w}{\tau_e \lambda w} \;,
\label{explkin}
\end{equation}
Hence, for sufficiently large $\tau_e$, $Y^{kin}$ approaches to the equilibrium value as $\tau_e^{-1}$. More details on the calculations are shown in Appendix B.

\section{V. Results}

In the following three subsections we present a systematic analysis of the FDR in Eq.~\eqref{FDR}.
Therein, we show the correlation function and the integrated response for different values of the parameters, and the corresponding relaxation timescales.
We also show the parametric plots $\chi_X(C_X)$, and we compare the $T_{FDR}$ obtained for two different bath temperatures, $T = 200,300$K. In subsection VA, we present the simulation results obtained for the end-to-end length $L$, in subsection VB we replicate the analysis for another observable, the total number of native contacts, $N_c$. In subsection VC, we calculate the kinetic temprature defined via Eq.~\eqref{kin}. 
Then, we discuss all the nonequilibrium temperatures which emerge from different definitions and observables, and we compare the results with the predictions of the 4-state model.

\subsection{A. Effective temperature for the end-to-end length}
To compute the effective temperature $T^L_{FDR}$, which is associated with the end-to-end length $L$, we evaluate numerically Eq.~\eqref{FDR}. The system is prepared in a NESS, at bath temperature $T$, with a pulling force $f = f_{bias} \pm \Delta f $, which switches with rate $1/\tau_e$. In such state, for $t_0 = 0$, the integrated correlation in Eq.~\eqref{corr} reads:
\begin{equation}
C_L(t) = \langle L^2(0) \rangle_{ss} - \langle L(0) L(t) \rangle_{ss} \; .
\label{eq:CL}
\end{equation}
At time $t_0$, a small steplike perturbation $\delta f \, \theta(t - t_0)$ in the force bias ($f_{bias} \to f_{bias} + \delta f$) is applied, shifting on average the end-to-end length by a quantity $\langle L(t) - L(0) \rangle$. Thus, from Eq.~\eqref{resp}, the integrated response function can be readily found:
\begin{equation}
\chi_L(t) = \frac{\langle L(t) - L(0) \rangle}{\delta f} \; .
\label{eq:RL}
\end{equation}
The response function is defined in the limit $\delta f \to 0$. We perform two sets of simulations for different values of the parameters, respectively $T = 300 K$, $f_{bias} = f_c = 15.3 $pN, $\delta f = 1$pN and $T=200 K$, $f_{bias} = f_c = 30$pN, $\delta f = 0.5$pN. Such values of the force bias correspond to the equilibrium folding-unfolding transition at the given temperatures (the former is also the experimental unfolding force at room temperature, the latter is predicted by our model, see also Fig.~\ref{fig1}). The chosen values of $\delta f$ are sufficiently small to prevent nonlinear contributions from significantly affecting the measure of $\chi_L(t)$. The amplitude of the time-dependent pulling force is $\Delta f = 10.0$pN for both the cases (the same value has been used to produce the time series described in subsection IIB). Finally, we span a large range of switching times, from $\tau_e = 10^2$ to $\tau_e = 10^4$. In Fig.~\ref{fig3}(a,d) we show $C(t)$ as a function of time. Note that, as $\tau_e$ increases, the asymptotic value of $C(t)$ becomes larger. Surprisingly, the susceptibility $\chi(t)$ decreases with $\tau_e$, which is apparently counterintuitive, see Fig.~\ref{fig3}(b,e). In fact, one would expect high-frequency external drives to lower the ability of the system to respond to external perturbations, as it acts to increase the disorder. Contrarily, in our system, the action of the switching force generates a significant raise of the susceptibility $\chi_\infty$ when $\tau_e$ becomes smaller. This is because the constant force bias $f_{bias}$ determines the direction of the molecule, which is chiefly oriented parallel to the direction of the force ($L$ is always positive at the transition, for $T=200$K and $T=300$K, see Fig.~\ref{fig1} and Fig.~\ref{fig2}). In other words, this results in a partial ordering of the native stretches, namely a prevalence of $\sigma_{ij} = +1$. Therefore, high frequency external drives aid the system to respond to external perturbations, similarly to what happens in the Ising model below the critical temperature, where the susceptibility increases as the temperature raises. This tendency is inverted when the RNA molecule is disordered, which occurs for very small forces and high temperatures, where $L \sim 0$ and the fraction of the positively--oriented stretches, $\sigma_{ij} = +1$, equals the fraction of the negatively--oriented ones, $\sigma_{ij} = -1$ (not shown).

\begin{figure*}[t]
\includegraphics[width=1.\textwidth]{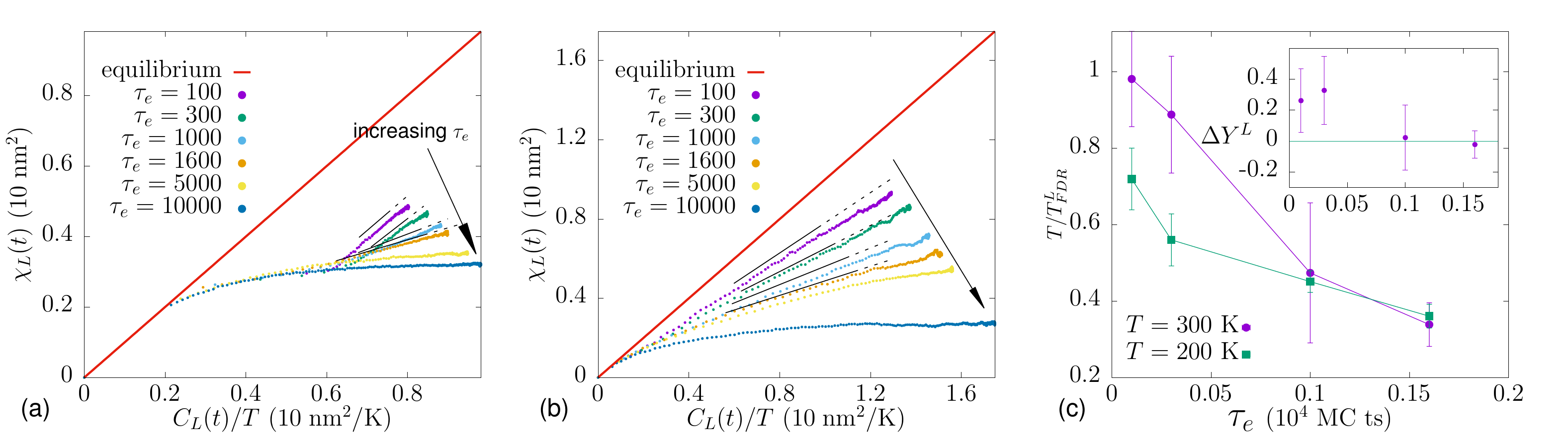}
\caption{\textbf{FDR and effective temperature for $L$.} Simulation were performed for $\tau_e = 100,300,1000,1600,5000,10000$. Parametric plot $\chi(C/T)$ for \textbf{(a)} $T=300 K$ and \textbf{(b)} $T = 200 K$. We observe a linear regime for $\tau_e \leq 1600$, with a slope which progressively lowers as $\tau_e$ increases. For $\tau_e = 5000$,$10000$ no linear regime is detected, corresponding to the out-of-equilibrium condition at which no $T_{FDR}$ emerges. This reflects in a nonvanishing curvature of the parametric plots (yellow and blue curves). Black solid lines were drawn to show the slope of the parametric curves in the linear regime, where the curvature is minimum (see Appendix A). Dashed lines are continuations of the fitting lines. \textbf{(c)} Violation parameter as a function of $\tau_e$, for the two values of bath temperature $T$. For $\tau_e \gtrsim 1000$ the ratio $T/T_{FDR}$ is the same for the two values of bath temperature $T$ (see Inset).}
\label{fig4}
\end{figure*}

The simple model in Section IV suggests that the correlation $C(t)$ and the response function $\chi(t)$ can be fitted by the following expressions:
\begin{equation}
\begin{split}
C(t) &\approx a_C+b_C(1-\text{e}^{-t/\tau^f_C})+c_C(1-\text{e}^{-t/\tau^s_C}),  \\
\chi(t) &\approx a_\chi+b_\chi(1-\text{e}^{-t/\tau^s_\chi }),
\end{split}
\label{eq:fit}
\end{equation} 

\begin{figure*}[t]
\includegraphics[width=1.\textwidth]{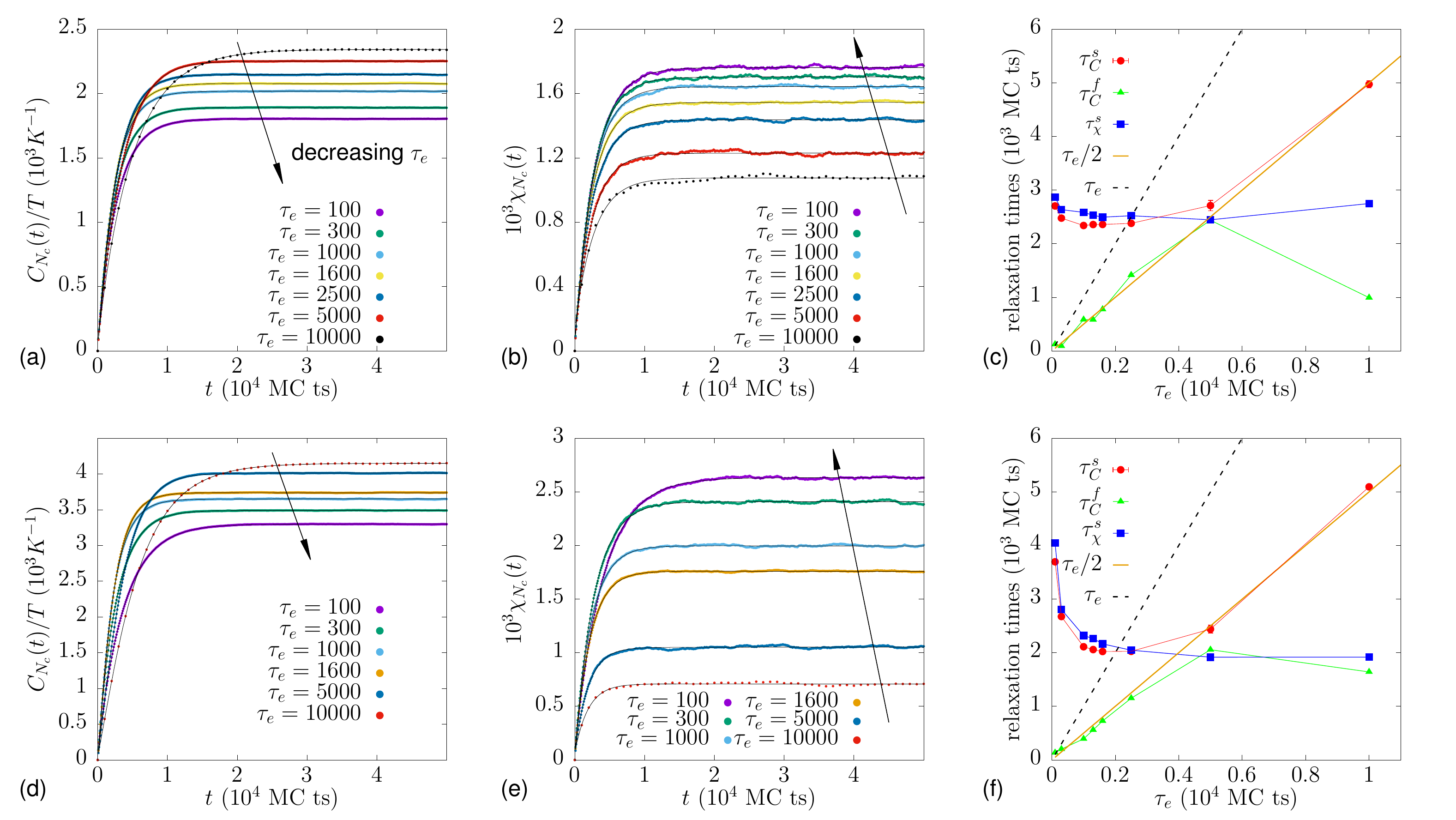}
\caption{\textbf{$\boldsymbol{C_{N_c}(t)}$, $\boldsymbol{\chi_{N_c}(t)}$ and relaxation timescales.} The integrated correlation $C_{N_c}(t)$ and response $\chi_{N_c}(t)$ curves for different $\tau_e$ are represented for two value of the bath temperature, \textbf{(a,b)} $T=300 K$ and \textbf{(d,e)} $T = 200 K$. The response curves are realized by applying a perturbation $\delta \epsilon = 0.02$ ($T = 300K$) and $\delta \epsilon = 0.01 $ ($T = 200K$) at $t = 0$. Two values of the perturbation were used to minimize nonlinear effects. As for the end-to-end length, the asymptotic values of correlation $C_\infty$ (respectively response $\chi_\infty$) increase (resp. decrease) as $\tau_e$ increases, especially in the $T=200K$ case. Relaxation timescales as a function of $\tau_e$ are shown for \textbf{(c)} $T=300$ and \textbf{(f)} $T=200K$. The plateau of $\tau^s_\chi$, is reached at $\tau_e \simeq 1000$ in both curves, but, unlike the $L$ case, the two curves differs substantially. Even though the starting points of the plateau are approximately the same ($\tau_e \simeq 1000$), their asymptotic values are consistently separated ($ \sim 2.5 \cdot 10^3 $ for $T=300 K$, panel (c), $\sim 2 \cdot 10^3$ for $T=200 K$, panel (f)).}
\label{fig5}
\end{figure*}

where $a_{C,\chi}$, $b_{C,\chi}$, $c_{C}$, $\tau^f_{C}$ and $\tau^{s,f}_{C,\chi}$ are fitting parameters, and the superscripts $f,s$ refers to `fast' and `slow' frequency mode. Note that the fitting expressions in Eq.~\eqref{eq:fit} differ from Eqs.~\eqref{Cmodes_approx} and \eqref{Rmodes_approx}. Indeed, fluctuations are not considered in the simple model described before; conversely, they are present in the full model, as shown in Fig~\ref{fig1}. They affect correlation and response functions in the very early times, and we take into account of such fluctuations by adding the constants $a_{C,\chi}$ to the expressions in Eqs.~\eqref{Cmodes_approx} and \eqref{Rmodes_approx}. We also remark that the response function does not decay with $\tau_e$ (as suggested by the simple 4-state model), and, therefore it can only be $\tau^s_\chi \equiv \tau_s$. 

We can now comment Figs.~\ref{fig3}(c,f), where we show the relaxation timescales as a function of the switching time $\tau_e$. Interestingly, the slowest timescale of the integrated correlation, $\tau^s_C$, is a nonmonotonical function of $\tau_e$ for both $T=200,300 K$.  This is not the case for the behavior of $\tau^s_\chi$, which seems to decrease monotonically towards the $\tau_e \to \infty$ equilibrium value.
Moreover, when there is a clear separation between the two relevant timescales, i.e. for $\tau_e \ll \tau_s$, we have that $\tau^s_\chi \simeq \tau^s_C \equiv \tau_s$, with a good overlap, especially for $T = 200$K. Correspondingly, the fast mode evolves with a typical timescale of $\tau_e/2$, as expected from the theory (in this case the fit is more accurate for $T=300$ K, see Fig.~\ref{fig3}(c)). We also observe that, for large $\tau_e$, the switching dynamics at long times takes over the relaxation dynamics of the perturbed system, which reflects in a substantial difference between $\tau^s_C$ and $\tau^s_\chi$. Indeed, for such values of $\tau_e$ the slowest relaxation timescale is $\tau_e/2$. We also recognize that this is the regime in which the violation parameter is time-dependent. 
The intersection between the $\tau^s_\chi$ and the $\tau_e$ line in Fig.~\ref{fig3}(c,f) separates approximately the regime in which the violation parameter $Y$ is constant from the regime in which $Y(t)$ is time-dependent. 

\begin{figure*}[t]
\includegraphics[width=1.\textwidth]{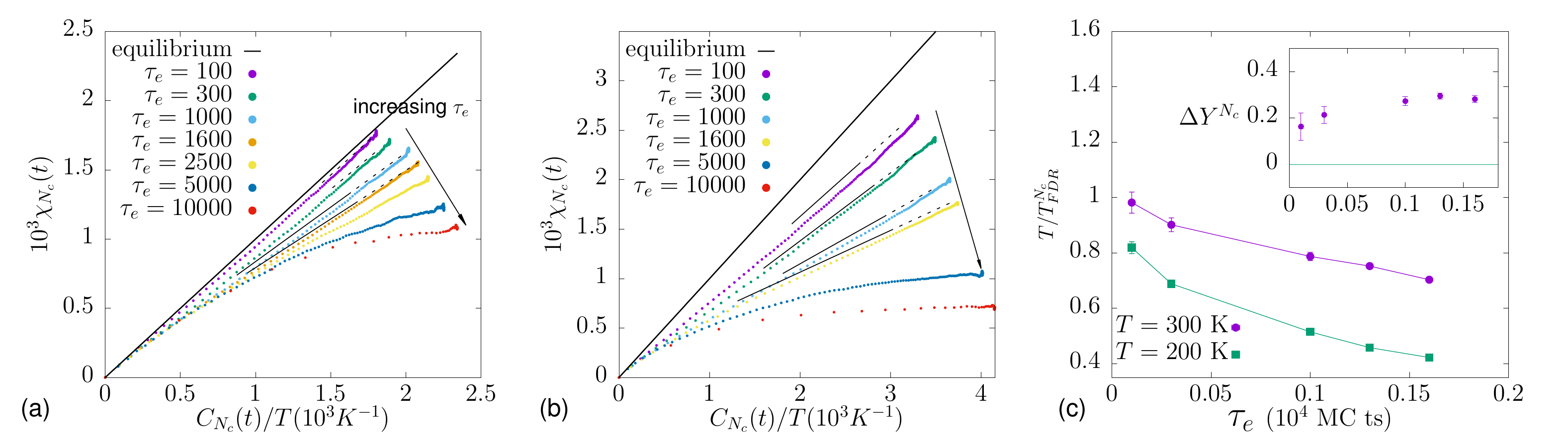}
\caption{\textbf{FDR and effective temperature for $N_c$.} Simulation were performed for different values of the switching time of the pulling force ($\tau_e = 100, 300, 1000, 1600, 2500, 5000, 10000$). Parametric plot $\chi(C/T)$ for \textbf{(a)} $T=300 K$ and \textbf{(b)} $T = 200 K$. Alike Fig.~\ref{fig4}, the fit in the minimum curvature interval (solid black lines) and their continuations (dashed black lines) are shown. \textbf{(c)} Unlike the $L$ case, here we have a marked difference in the behavior of $T/T^{N_c}_{FDR}$ for the two values of bath temperature $T$ used. The discrepancy is more relevant for large $\tau_e$, since the FDR temperature departs significantly from the equilibrium value. In the inset we show the discrepancy $\Delta Y^{N_c}$, which remains approximately constant for each value of $\tau_e \gtrsim 1000$.}
\label{fig6}
\end{figure*}

When $\tau_e < \tau_s$, a linear regime for large $t$ emerges in the parametric plot in Fig.~\ref{fig4}(a,b). Such behavior signals the appearance of an effective temperature, $T^L_{FDR}$, which deviates from the one of the bath. In particular, we always find $T_{FDR} > T$, which means that the activity of the system effectively converts to thermal-like fluctuations~\cite{Baiesi2009,Ritort2015,Cugliandolo2011}. The effective temperature appears after a certain time, $\tau_c$, which is consistent with previous predictions and experiments on several systems~\cite{Ritort2015}. The existence of such time $\tau_c$ is ascertained by the presence of an initial transient where $T \partial \chi/ \partial C \simeq 1$, see also Fig.~\ref{figA} in the Appendix A. The occurrence of a transient time is due to the violation of the FDR introduced by the `fast' mode $2/\tau_e$ in the correlation $C(t)$, and is very pronounced in the $T = 300 K$ parametric plot in Fig.~\ref{fig4}(a). Conversely, when $\tau_e > \tau_s $, namely when the force timescale is larger then the intrinsic relaxation timescale, no linear regime emerges, and the effective temperature cannot be defined. We remark that the behavior of the parametric plot in Fig~\ref{fig4}(a) differs from the one shown in Fig.~\ref{fig4}(b) (and from the other cases shown in the following subsecton). A `plateau' in the parametric plot appears at intermediate values of $C(t)$. It should be noted that such values correspond to a timescale of the order of $\tau_c \approx \tau_e/2$, whereas in this paper we focus on the definition of $T_{FDR}$ which appear at times much larger than $\tau_e$. Even though this behavior might reveal some interesting physics at these intermediate timescales, we will address the investigation of such behaviour to future works. In Fig.~\ref{fig4}(c) we show the violation parameter $Y^L(T) = T/T^L_{FDR}$, see also Eq.~\eqref{efftemp}, obtained by a linear fit of the curves in the parametric plots in Figs.~\ref{fig4}(a,b), as a function of $\tau_e$, for $T = 300 K$ and $T = 200 K$. Note that the effective temperature $T^L_{FDR}$ is always an increasing function of $\tau_e$. Interestingly, increasing the switching time $\tau_e$ decreases the deviation
\begin{equation}
\Delta Y^L = Y^L(T_1) - Y^L(T_2) \; ,
\end{equation}
where $T_1 = 300$ and $T_2 = 200$, see inset in Fig.~\ref{fig4}. Remarkably, for $\tau_e \gtrsim 1000$ the curves overlap, accordingly with the statistical error of the fit, as shown in the inset of Fig.~\ref{fig4}(c).

\subsection{B. Effective temperature for the total number of native contacts}

Several lines of evidence point to the fact that the effective temperature defined via a fluctuation-dissipation relation is dependent on the particular observable~\cite{Droz2009,Levis2015}. However, it seems that especially for systems with slow relaxation and aging, such differences in the effective temperatures tend to disappear \cite{Cugliandolo1997}. Various results suggest that, when a very slow relaxation timescale governs the long-time dynamics of the system, every frequency-dependent observable show the same parametric plot, and, then, the same $T_{FDR}$ \cite{Levis2015,Cugliandolo1997}. Here, we question whether some of these properties are present in our folding-unfolding RNA dynamics. We replicate the same analysis in subsection VA for another variable which describes our system, the number of native contacts $N_c$. Such choice is natural, since in the Hamiltonian in Eq.~\eqref{hamiltonian} $N_c$ is already coupled with its conjugate intensive variable, $-\epsilon$, which represents the energetic gain of a single atomic contact between two residues, when in their native configuration. 

Therefore, in order to compute the effective temperature $T^{N_c}_{FDR}$ for a given set of the parameters, we prepare the NESS with the same protocol used before (by driving the system out of equibrium via a switching force $f_{bias} \pm \Delta f$). Then, we perturb the system at time $t_0$ by increasing the value of $\epsilon$ by a small quantity $\delta \epsilon $ ($\delta \epsilon = 0.01,0.02$ at $T=200$,$300K$ respectively). Alike in Section VA, we evaluate the integrated correlation and response functions:
\begin{equation}
C_{N_c}(t) = \langle N_c^2(0) \rangle_{ss} - \langle N_c(0) N_c(t) \rangle_{ss} \; ,
\label{eq:CNc}
\end{equation}
\begin{equation}
\chi_{N_c}(t) = \frac{\langle N_c(t) - N_c(0) \rangle}{\delta \epsilon} \; .
\label{eq:RNc}
\end{equation}
We range the force switching time $\tau_e$ from $10^2$ to $10^4$ Monte Carlo time steps. The results are qualitatively the same: increasing $\tau_e$ produces an increase of the long-time integrated correlation $C(t)$, as well as a decrease of the susceptibility $\chi_\infty$, see Figs.~\ref{fig5}(a,b,d,e). The corresponding `long' relaxation timescale $\tau^s_\chi$ displays the same seemingly monotonic behavior as in the end-to-end length case.
Differently from the previous case, the asymptotic value of $\tau^s_\chi$ (large $\tau_e$) varies with the bath temperature $T$, as shown in Figs.~\ref{fig5} (c,f). Thus, there is a strong dependence on $T$ of the relaxation properties of the observable $N_c$, even for large $\tau_e$. However, the general features of the nonequilibrium correlation and response functions also hold for this variable, that is $\tau^f_C \simeq \tau_e/2$ for small $\tau_e$, $\tau^s_C \simeq \tau_e/2$ for large $\tau_e$.

In Fig.~\ref{fig6} it can be seen that the parametric plots deviates from the equilibrium line ($\chi_{N_c} = C_{N_c}(T)/T$) much more in the $T = 200 K$ case than in the $T = 300 K$ case. Moreover, for $T = 300 K$ the region of violation of the FDR in Eq.~\eqref{FDR} (nonzero curvature) is much less pronounced here than in the cases shown in Fig.~\ref{fig4}. This is due to the reduced fluctuations in the basins associated with the folding/ordered and unfolding/disordered states, compare Fig.~\ref{fig2}(e) with Fig.~\ref{fig2}(f). In fact, such difference is much less marked when the extents of fluctuations into the two basins resemble each other, as in the $T = 200 K$ case (not shown). Nonetheless, a region of thermal-like behavior of fluctuations emerges for both $T$, validating the generality of the hypothesis made in Section III. Indeed, a linear regime $T^{N_c}_{FDR}$ arises for sufficiently small $\tau_e$. As $\tau_e$ increases, the linear trend starts at larger times ($\tau_c $ increases), enlarging the violation region, until, for large enough switching times ($\tau_e \gtrsim 2500$), the whole parametric plot displays a nonzero curvature (see also Appendix A). The strong dependence of the nonequilibrium slow relaxation timescale is more evident in the effective temperature $T^{N_c}_{FDR}$. In Fig.~\ref{fig6}(c) we show the violation parameter as a function of $\tau_e$, for both the bath temperatures. Albeit the qualitative behavior is similar to the one in Fig.~\ref{fig4}(c) for the effective temperature $T^L_{FDR}$, here the deviation between the two curves is statistically significant, as shown in the inset of Fig.~\ref{fig6}(c). There, we can see how the difference $\Delta Y^{N_c}=T_1/T^{N_c}_{1,FDR} - T_2/T^{N_c}_{2,FDR}$, with $T_1 = 300$ and $T_2 = 200$ is constantly nonzero in the entire range of switching times. 

\subsection{C. Comparison between FDR and kinetic temperature}

In this subsection, we evaluate a `kinetic' temperature, which is calculated from the rate of heat exchanged by the system and a second thermal bath. It is still useful to compute the kinetic temperature associated to different variables, as already done in the previous subsections. In light of this, we will consider the native stretches orientations $\sigma_{ij}$ and the nativeness of the RNA bases $m_i$. To calculate the kinetic temperature for the variables $\sigma \equiv \{\sigma_{ij}\}$ (respectively $m \equiv \{m_i\}$) indipendently, which we will denote with $T^\sigma_{kin}$ (respectively $T^m_{kin}$), we apply the procedure in Section IIIB by computing the variation in energy (step (ii)) while $m_i$ (respectively $\sigma_{ij}$) is constant. For instance, in order to find $T^\sigma_{kin}$, we only consider the contribution $E(\sigma_{t+1},m_t,f(t))-E(\sigma_{t},m_t,f(t))$ to the total heat exchanged at time $t+1$.

\begin{figure}[t]
\includegraphics[width=.48\textwidth]{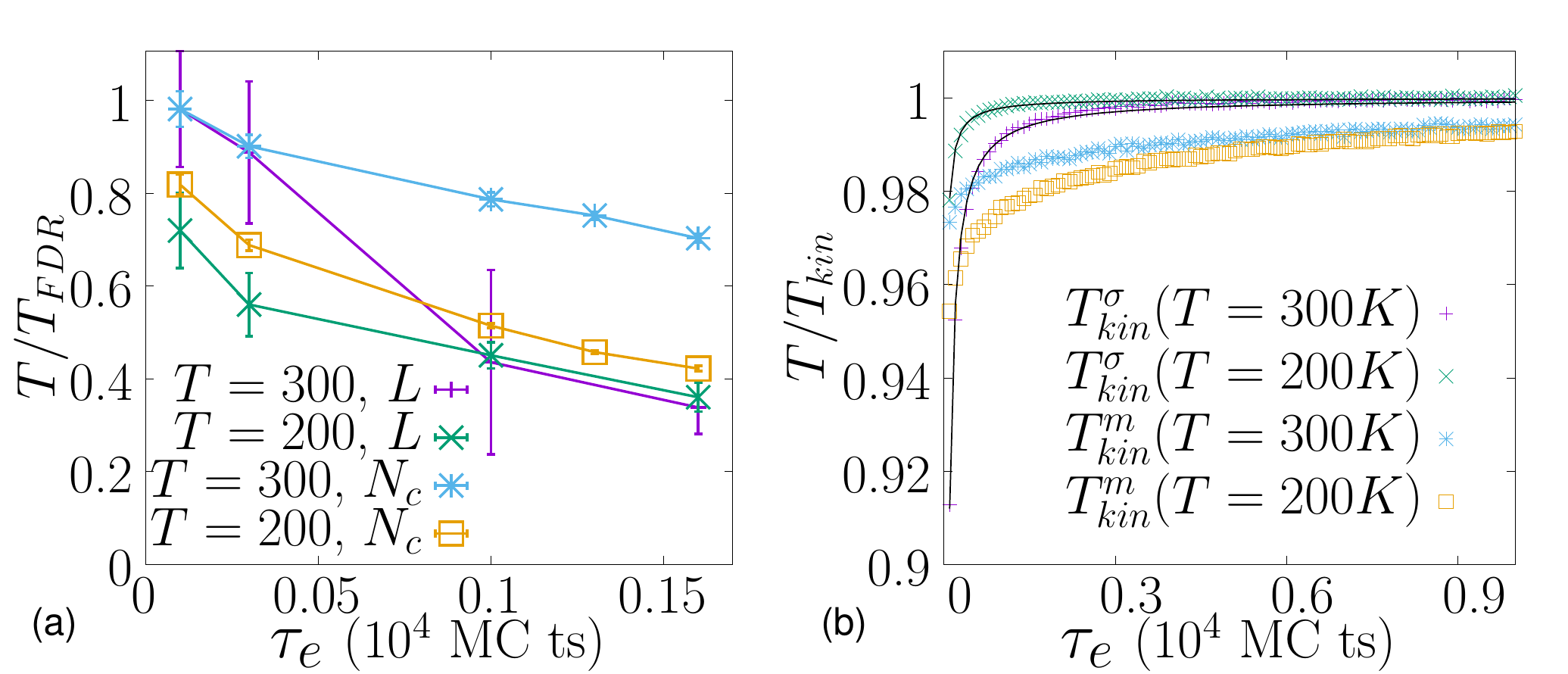}
\caption{\textbf{Effective and kinetic temperatures.} The violation parameter relative to the FDR and the kinetic temperatures are here plotted, as a function of $\tau_e$. \textbf{(a)} Comparison of the violation parameters for different values of the parameters, shown also in Figs.~\ref{fig4}(c) and \ref{fig6}(c). Note that there is no apparent match between the values of the violation parameter defined via FDR, except for large $\tau_e$, when $T=200,300 K$, in the $L$ case. \textbf{(b)} The ratio $T/T_{kin}$ is shown in this panel, as a function of $\tau_e$, for both the variables $\sigma$ and $m$. In this case, $T_{kin}$ displays a totally different behavior from the FDR temperature, as it decreases with $\tau_e$. Observe also that the deviations from the bath temperature $T$ are small, even for high-frequency drives. Solid lines represent the functional behavior predicted by the 4-state model.}
\label{fig7}
\end{figure}

In Fig.~\ref{fig7} we show a comparison amongst all the effective temperatures computed via FDR and the kinetic temperature calculated from the exchanged heat, for different values of $\tau_e$ and $T$. In Fig.~\ref{fig7}(a) we restrict to the effective temperatures $T_{FDR}$ calculated in Section V. Note that the close-to-equilibrium condition here is represented by small $\tau_e$. For such values an effective temperature can always be defined by FDR, but is very close to the bath temperature $T$ (the FDR temperature is closer to the equilibrium temperature for $T=300K$). For larger values of $\tau_e$, there is no apparent collapse in the effective temperature curves at the same bath temperature, whereas in the case of $T^L_{FDR}$, when $\tau_e \gtrsim 1000$ the two curves overlap within the error bars. Interestingly, this occurs when the effective temperature $T^\sigma_{kin}$ measured by the `thermometer' is almost equal to $T$, as shown in Fig~\ref{fig7}(b).

In Fig.~\ref{fig7}(b), the effective temperature $T_{kin}$ is represented as a function of $\tau_e$, obtained for $m$ and $\sigma$ with the abovementioned precedure. Deviations from the bath temperature are signaled for small $\tau_e$. By increasing $\tau_e$, the effective temperature approaches $T$. Note that such procedure allows to define an effective temperature for any value of the driving switching time, and that, differently from the FDR, the close-to-equilibrium condition is for large values of $\tau_e$. For large switching times, a small departure of $T^m_{kin}$ from the bath temperature is detected, while the deviation of the kinetic temperature $T^\sigma_{kin}$ from the bath temperature $T$ is almost null for both $T=300 K$ and $T=200 K$, and large $\tau_e$, as the ratio $T/T^{\sigma}_{kin} \simeq 1$. Additionally, we also provide a direct comparison between $T^\sigma_{kin}(\tau_e)$ obtained from our simulations and the behavior predicted in Eq.~\eqref{explkin}. We find that the function $1-a/\tau_e$ (being $a$ a fitting parameter) fits perfectly the curves in Fig.~\ref{fig7}(b) for the variable $\sigma_{ij}$, which is conjugated to the amplitude of the force in the hamiltonian (see Eqs.~\eqref{L} and~\eqref{hamiltonian}). This results assesses the validity of the 4-state model for large $\tau_e$.

\section{VI. Discussion and conclusions}

In this paper we explore the possibility of extending the notion of temperature in a nonequilibrium context, for a biological system similar to the one experimentally studied in Ref.~\cite{Ritort2015}. We consider a model for the RNA/protein folding whose equilibrium properties have been widely studied in previous works. Using this model, we offer an extensive description of the nonequilibrium properties of the PG5A RNA hairpin, and, therein, we focus on the emergence of different effective temperatures related to several variables. We perform our measures in the nonequilibrium steady state, or NESS, which is prepared by forcing the molecule by means of an external random switching force of switching time $\tau_e$.

Firstly, we study the FDR in Eq.~\eqref{FDR}, by computing numerically correlation and response function. We perform the same procedure for two different observables, the end-to-end length of the molecule, $L$, and the number of native contacts, $N_c$.
The analysis of the FDR produces results qualitatively in accordance with those in~\cite{Ritort2015}, for both the observables. Two timescales are relevant, the intrinsic relaxation timescale $\tau_s$ and the switching time of the drive $\tau_e$.
One finds that when $\tau_s > \tau_e$, an effective temperature can be defined.
In this regime, a linear trend in the correlation-response parametric plot appears at large times. In particular, the violation parameter $Y$ decreases as $\tau_e$ increases, and, in the same NESS, assumes different values for different observables. This dependence has also been proved analytically in previous works~\cite{Droz2009}. 

Secondly, we propose a different temperature definition, which takes into account the mean instantaneous heat exchanged with another weakly-interacting bath. Differently from the FDR effective temperature, the `kinetic' temperature defined below \textit{(i)} reflects the instantaneous dissipative properties of the system and \textit{(ii)} is related to the change of the values of some microscopic variables of the model. 
This procedure mimics the equilibrium prescription of measuring temperatures by using a `thermometer'.
The temperature at which this bath does not exchange energy with the system, is defined as the \textit{kinetic temperature}, which can be treated as another nonequilibrium characteristic of the system.
Interestingly, such kinetic temperature is well-defined for every force switching timescale $\tau_e$, is higher for small $\tau_e$ and tends to the equilibrium temperature for large $\tau_e$.

We should also remark that both $T_{kin}$ and $T_{FDR}$ display the same behavior as a function of the drive amplitude, as they grow when $\Delta f$ increases. One can find, by using the simple $4$-state model described in Section~VA, that $T_{FDR,kin} - T \propto \Delta f^2 + \mathcal{O}(\Delta f^3)$. Simulations on the full model are in accordance with this prediction (not shown). Therefore, it can be assessed that both the effective temperature $T_{FDR}$ and $T_{kin}$ measure how far the system is from equilibrium. Nonetheless, the behavior of the two temperature, and the related violation parameters, with respect to the frequency of the drive is opposite. We also verified that the heat exchanged between the hairpin and the thermal bath at temperature $T$ is proportional to the difference $T_{kin} - T$ (not shown), as expected for two systems at different but similar temperatures which are kept in contact. Additionally, in the $4$--state model, under the condition $M_{31} = M_{24}$, at fixed $x$, one can find that
\begin{equation}
\frac{P^0_{(s_+,x)}}{P^0_{(s_-,x)}} = \exp(-(E(s_+,x)-E(s_-,x))/T_{kin}) \; ,
\end{equation}
confirming that $T_{kin}$ is a possible measure of nonequilibrium temperature in the steady state. We address to future work further assessments of the robustness of the `kinetic' temperature. In this regards, a study of the fluctuations of the exchanged energy $E(\Sigma')-E(\Sigma)$ might be significant.

In Ref.~\cite{Petrelli2020} the conceptual difference between a FDR effective temperature and a `kinetic' temperature (defined straightforwardly via the kinetic energy) has been explored in the context of active matter. Here, we find that the two effective temperatures are intrinsically different, as they capture different features of the nonequilibrium dynamics. The FDR describes the long time-delay thermodynamic behavior of a nonequilibrium system; if FDR hold, then the system respond equally to both a small ``external'' perturbation and to an ``internal'' perturbation (or fluctuation), similarly to what happens at equilibrium.
Thus, we found that the appearance of an effective temperature $T_{FDR}$ is strictly connected to the long relaxation timescale.
Conversely, the kinetic temperature is more related to the instantaneous thermodynamic properties, which can mainly inform of the frequency of the time-dependent external drive.

As a conclusion, we observe that, in previous works, a theoretical framework on the linear response for system out of equilibrium has been developed. There, the connection between the time-symmetric contribution to the linear response, also called \textit{frenesy}~\cite{Baiesi2009}, and the effective temperature has been established~\cite{Lippiello2005}. In nonequilibrium conditions, the integrated FDR reads $\chi(t) = (C(t) + K(t))/2$, where $C(t)$ is an equilibrium-like correlation, while $K(t)$ has a frenetic (time-symmetric) origin, which reduces to $C(t)$ in the equilibrium limit. This latter is an exclusive nonequilibrium contribution; it would be interesting to calculate such terms in our model, both analytically and numerically, evidencing their dependence on the relevant parameters, and work is in progress along these lines.

\begin{table*}[ht]
  \begin{center}
    \caption{\textbf{fitting parameters.}}
    \label{tab:table1}
    \begin{tabular}{c|c|c|c|c|c|c|c} 
      $T$ & $X$ & $\tau_e$ & $a_{C}$ & $b_{C}$ & $c_{C}$ & $a_{\chi}$ & $b_{\chi}$ \\
      \hline
     $ 300 $ & $L$ & $ 100 $ & $ 31.9 \pm 2.0 $ & $ 20.3 \pm 0.02 $ & $ 27.8 \pm 2.0 $ & $ 19.7 \pm 0.07 $ & $ 28.5 \pm 0.08 $ \\
$ 300 $ & $L$ & $ 300 $ & $ 37.6 \pm 0.2 $ & $ 23.2 \pm 0.04 $ & $ 23.9 \pm 0.2 $ & $ 19.0 \pm 0.09 $ & $ 27.3 \pm 0.09 $ \\
$ 300 $ & $L$ & $ 1000 $ & $ 33.6 \pm 0.1 $ & $ 29.7 \pm 0.1 $ & $ 24.7 \pm 0.08 $ & $ 16.7 \pm 0.1 $ & $ 26.3 \pm 0.1 $ \\
$ 300 $ & $L$ & $ 1600 $ & $ 31.2 \pm 0.3 $ & $ 34.2 \pm 0.3 $ & $ 24.2 \pm 0.08 $ & $ 15.9 \pm 0.1 $ & $ 25.0 \pm 0.1 $ \\
$ 300 $ & $L$ & $ 5000 $ & $ -193 \pm 1302 $ & $ 266 \pm 1302 $ & $ 21.9 \pm 0.03 $ & $ 12.8 \pm 0.1$ & $ 22.3 \pm 0.1 $ \\
$ 300 $ & $L$ & $ 10000 $ & $ -119 \pm 245 $ & $ 196 \pm 245 $ & $ 20.7 \pm 0.05 $ & $ 11.3 \pm 0.09 $ & $ 20.7 \pm 0.09 $ \\
$ 200 $ & $L$ & $ 100 $ & $ 4.2 \pm 0.8 $ & $ 116.5 \pm 0.07 $ & $ 8.9 \pm 0.8 $ & $ 83.2 \pm 0.1 $ & $ 9.4 \pm 0.2 $ \\
$ 200 $ & $L$ & $ 300 $ & $ 5.7 \pm 0.5 $ & $ 125.2 \pm 0.1 $ & $ 6.2 \pm 0.5 $ & $ 76.0 \pm 0.2 $ & $ 9.9 \pm 0.3 $ \\
$ 200 $ & $L$ & $ 1000 $ & $ -24.2 \pm 4.4 $ & $ 164 \pm 4 $ & $ 5.3 \pm 0.2 $ & $ 64.6 \pm 0.2 $ & $ 6.6 \pm 0.3 $ \\
$ 200 $ & $L$ & $ 1600 $ & $ -18.6 \pm 0.9 $ & $ 163 \pm 1 $ & $ 6.2 \pm 0.4 $ & $ 55.7 \pm 0.3 $ & $ 6.6 \pm 0.3 $ \\
$ 200 $ & $L$ & $ 5000 $ & $ -95 \pm 5 $ & $ 246 \pm 6 $ & $ 5.3 \pm 0.1 $ & $ 49.1 \pm 0.2 $ & $ 5.4 \pm 0.2 $ \\
$ 200 $ & $L$ & $ 10000 $ & $ -45.1 \pm 0.8 $ & $ 215.1 \pm 0.9 $ & $ 4.4 \pm 0.1 $ & $ 22.0\pm 0.2 $ & $ 5.1 \pm 0.2 $ \\
$ 300 $ & $N_c$ & $ 100 $ & $ 33 \pm 4 $ & $ 1761.2 \pm 0.4894 $ & $ 9 \pm 45 $ & $ 1713 \pm 2 $ & $ 49 \pm 2 $ \\
$ 300 $ & $N_c$ & $ 300 $ & $ 27.1 \pm 8 $ & $ 1861.3 \pm 0.5012 $ & $ 2 \pm 8 $ & $ 1667 \pm 2 $ & $ 34 \pm 2 $ \\
$ 300 $ & $N_c$ & $ 1000 $ & $ -58 \pm 3 $ & $ 2057 \pm 3 $ & $ 19 \pm 1 $ & $ 1598 \pm 2 $ & $ 44 \pm 2 $ \\
$ 300 $ & $N_c$ & $ 1600 $ & $ -167 \pm 8 $ & $ 2221 \pm 8 $ & $ 23 \pm 1 $ & $ 1503 \pm 2 $ & $ 41 \pm 2 $ \\
$ 300 $ & $N_c$ & $ 2500 $ & $ -782 \pm 84 $ & $ 2903 \pm 85 $ & $ 26 \pm 1 $ & $ 1382 \pm 2 $ & $ 54\pm 2 $ \\
$ 300 $ & $N_c$ & $ 5000 $ & $ -9353 \pm 23220 $ & $ 11565 \pm 23220 $ & $ 40.1 \pm 0.9 $ & $ 1163 \pm 4 $ & $ 65 \pm 4 $ \\
$ 300 $ & $N_c$ & $ 10000 $ & $ 56 \pm 12 $ & $ 2232.47 \pm 8 $ & $ 53 \pm 17 $ & $ 935 \pm 24 $ & $ 139 \pm 24 $ \\
$ 200 $ & $N_c$ & $ 100 $ & $ 91 \pm 7 $ & $ 3208.1 \pm 0.8 $ & $ -0.6 \pm 8.1 $ & $ 2550 \pm 2 $ & $ 81 \pm 2 $ \\
$ 200 $ & $N_c$ & $ 300 $ & $ 58 \pm 4 $ & $ 3432 \pm 1 $ & $ -0.5 \pm 4 $ & $ 2366 \pm 4 $ & $ 41 \pm 4 $ \\
$ 200 $ & $N_c$ & $ 1000 $ & $ -225 \pm 6 $ & $ 3862 \pm 6 $ & $ 11 \pm 5 $ & $ 1962 \pm 3 $ & $ 34 \pm 3 $ \\
$ 200 $ & $N_c$ & $ 1600 $ & $ -444 \pm 8 $ & $ 4129 \pm 9 $ & $ 11 \pm 3 $ & $ 1733 \pm 1 $ & $ 26 \pm 1 $ \\
$ 200 $ & $N_c$ & $ 2500 $ & $ -2340 \pm 77 $ & $ 6174 \pm 78 $ & $ 8.96015 \pm 1 $ & $ 1411 \pm 4 $ & $ 49 \pm 4 $ \\
$ 200 $ & $N_c$ & $ 5000 $ & $ -13163 \pm 5540 $ & $ 17165 \pm 5541 $ & $ 11 \pm 2 $ & $ 998 \pm 5 $ & $ 51 \pm 5 $ \\
$ 200 $ & $N_c$ & $ 10000 $ & $ -762 \pm 30 $ & $ 4862 \pm 38 $ & $ 48.448 \pm 14 $ & $ 641 \pm 27 $ & $ 67 \pm 27 $ \\

    \end{tabular}
  \end{center}
\end{table*}

\begin{figure}[t]
\includegraphics[width=.5\textwidth]{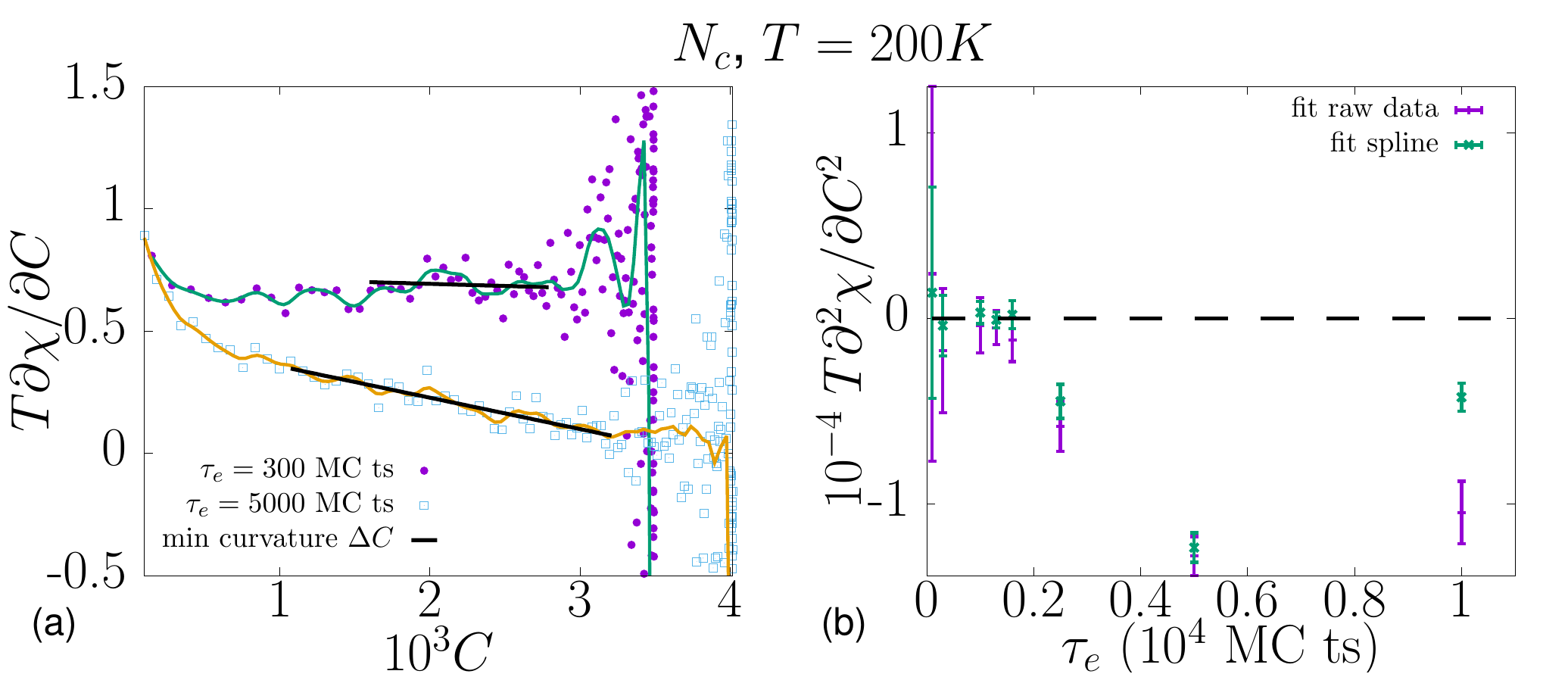}
\caption{\textbf{Violation parameter and curvature of $\chi(C)$.} In this figure, the first and second derivative of the parametric plot are shown, in the representative case of $N_c$ at $T=200 K$. \textbf{(a)} Purple and cyan points represent the violation parameter of the parametric plots in Fig.~\ref{fig6} at each sampled point $C(t_i)$, for $\tau_e = 300$ and $\tau_e = 5000$ respectively. Note that the violation parameter is close to $1$ at small $C$, and decreases to smaller values as $C$ increases. Solid colored lines are the interpolating weighted splines. For $\tau_e = 300$ the violation parameter becomes approximately horizontal, signalling the presence of a thermal-like regime, whilst for $\tau_e = 5000$, the violation parameter changes linearly with $C$. The fits of the minimum curvature interval are shown (black solid lines), as well as the threshold used to cut off the noisy region at large $C$ \textbf{(b)} The curvature (i.e. the slope of the linear fit in panel (a)) obtained from the raw data (purple points) and the interpolated points (green points). For $\tau_e \lesssim 1600$ the curvature is approximately null, as expected in a thermal-like regime.}
 
\label{figA}
\end{figure}

\section{Appendix A: CURVATURE AND FITTING PROCEDURE}
In this Appendix we describe the fitting procedure used to extract the violation parameter $Y(t)$ shown in Fig.~\ref{fig4}(c) and Fig.~\ref{fig6}(c). In order to find the crossover between the linear ($Y(t)$ is constant) and the nonlinear ($Y(t)$ is time-dependent) regimes, we compute $T \partial \chi(t)/ \partial C(t)$, for both the observables $L$ and $N_c$, for any $\tau_e$ used in our simulations. The first derivative is calculated with a simple forward difference scheme:
\begin{equation} 
\frac{\partial \chi(t)}{ \partial C(t)} \approx \frac{\chi(t_{i+1})-\chi(t_{i})}{C(t_{i+1})-C(t_{i})}.
\end{equation}
where $t_{i,i+1}$ are two subsequent sampled times. If the parametric plot is linear, the first derivative of the parametric plot should be horizontal, which corresponds to a null curvature ($T \partial^2 \chi(t)/ \partial^2 C(t) = 0$).

In Fig.~\ref{figA}(a), we show a representative plot of the $T \partial \chi(t)/ \partial C(t)$ as a function of $C(t)$, for the simulations performed for the observable $L$ at $T = 200$K; in order to filter out the noise, we also show a weighted spline which interpolates the points (solid colored curves) as a guide to the eye. From the simple $4$-state model (see main text), the linear region of the parametric plot should appear for $t \gtrsim \tau_c$, being $\tau_c$ a certain critical time scale of the order of $\tau_e$. Hence, we search for the interval with the minimum $\partial^2 \chi(t)/ \partial C(t)^2$ within $\left[ C(\tau_e/2),C^* \right]$, where $C^*$ is a threshold value chosen arbitrarly to exclude the noisy region for large $C$ (vertical dashed lines). Then, we perform a linear fit of both the raw data and the interpolating points in the selected interval; the slope of the fitting line represents the curvature $T \partial^2 \chi(t)/ \partial^2 C(t)$. Note that for $\tau_e = 300$, the fitting line is approximately horizontal, or, in other words, an effective temperature can be defined. Conversely, for $\tau_e = 5000$ the fit produces a nonnull slope, which means that no effective temperature can be detected. In Fig.~\ref{figA}(b), the curvatures for each value of $\tau_e$ are plotted; a clear crossover between a noncurve regime and a regime in which the curvature is nonnzero appear, signalling the upper limit of the range of $\tau_e$ in which $T_{FDR}$ can be defined. We should remark that such procedure gives robust results for any case studied in the paper.

The violation parameter $Y(t)$ has been found by fitting the $T \partial \chi(t)/ \partial C(t)$ plots (raw data) with the function $f(C) = Y$, in the interval of minimal curvature. The obtained values of $Y$ has been reported in Fig.~\ref{fig4}(c) and Fig.~\ref{fig6}(c), along with the relative statistical error on the fit. For completeness, in Table~\ref{tab:table1} we report the values of the parameters (and the relative statistical errors) in Eq.~\eqref{eq:fit} obtained by fitting the curves in Fig.~\ref{fig3}(a,b,d,e) and Fig.~\ref{fig5}(a,b,d,e), using a nonlinear least-squares Marquardt-Levenberg algorithm~\cite{mardquardt}.

\section{Appendix B: KINETIC TEMPERATURE IN THE 4-STATE MODEL}

In this Appendix we aim to calculate the kinetic temperature for the 4-state model described in Section V. In particular, we solve Eq.~\eqref{kin} to find an explicit expression of the rate of absorbed heat, then we set $\langle \dot{Q}_X \rangle = 0$ to find the kinetic temperature. In fact, as detailed in the main text, we imagine that our system can exchange heat with a second weakly-interacting thermal bath at temperature $T_{th}$. Hence, the kinetic temperature of the system is the temperature $T_{kin}=T_{th}^* $ at which it corresponds a vanishing flow of energy between the system and the second bath; in other words, when $T_{kin} = T^*_{th}$ one finds $\langle \dot{Q}_X \rangle = 0$ . From Eq.~\eqref{matrix}, one can evaluate the stationary probability distribution $\mathbf{P}^0 = \left( P^0_{(s_+,x_+)}  \; P^0_{(s_+,x_-)} \; P^0_{(s_-,x_+)} \; P^0_{(s_-,x_-)}\right)$ as the eigenvector of $\mathbb{M}$ associated with the null eigenvalue ($\mu_0 = 0$). A direct calculation gives:
\begin{equation}
\begin{split}
P^0_{(s_+,x_+)} &= \Gamma \left[M_{13} (M_{24}+M_{42}) + \frac{(M_{13}+M_{24})}{\tau_e} \right] \; ,\\ 
P^0_{(s_+,x_-)} &= \Gamma \left[M_{24}(M_{13}+M_{31}) + \frac{(M_{13}+M_{24})}{\tau_e} \right] \; ,\\ 
P^0_{(s_-,x_+)} &= \Gamma \left[M_{31}(M_{24}+M_{42}) + \frac{(M_{31}+M_{42})}{\tau_e} \right] \; ,\\ 
P^0_{(s_-,x_-)} &= \Gamma \left[M_{42}(M_{13}+M_{31}) + \frac{(M_{31}+M_{42})}{\tau_e} \right] \; ,\\ 
\label{expleigen}
\end{split}
\end{equation}
where 
\begin{equation}
\Gamma = \frac{\tau_e}{2\tau_e(M_{13}+M_{31})(M_{24}+M_{42}) + \sum M}
\end{equation}
is a normalization constant which ensures that $\sum P^0_{(s,x)} = 1$, with $\sum M = M_{13} + M_{31} + M_{24} + M_{42}$; then, if $s = s_{\pm}$ and $x = \pm \Delta x$ are conjugated in the hamiltonian (being $\Delta x > 0$ and $s_+ - s_- > 0$), the energy change at time $t$ in the presence of the external drive $x$ is $E(s_{t+1},x)-E(s_{t},x) = -x(s_{t+1} - s_{t})$. Thus, if we define $\lambda_{th}$ the intrinsic rate of exchange of heat between the system and the second bath, Eq.~\eqref{kin} reads:
\begin{equation}
\begin{split}
\langle \dot{Q}_X \rangle &= \lambda_{th} \Delta x (s_+ - s_-) \biggl\{ \left[ P^0_{(s_+,x_+)}  + P^0_{(s_-,x_-)}\right] w_{th} \\
&- \left[ P^0_{(s_-,x_+)}  + P^0_{(s_+,x_-)}\right] \biggl\} ,
\label{explkinS}
\end{split}
\end{equation}
where $w_{th} \equiv \mathrm{exp} \{ -\beta_{th} \Delta x (s_+ - s_-) \}$, with $\beta_{th} = 1/T_{th}$.
From Eq.~\eqref{explkinS}, one can find the kinetic violation parameter as a function of the bath temperature $T$, the amplitude $\Delta x$ of the switching external drive and the transition rates between the $4$ states of the model. By setting $\langle \dot{Q}_X \rangle = 0$, it follows:
\begin{equation}
\frac{T}{T_{kin}} = \frac{T}{\Delta x (s_+ - s_-)} \mathrm{ln} \left[\frac{P^0_{(s_+,x_+)}  + P^0_{(s_-,x_-)}}{P^0_{(s_-,x_+)}  + P^0_{(s_+,x_-)}}\right] \; .
\label{genkin}
\end{equation}
By putting the expressions in Eq.~\eqref{expleigen} into Eq.~\eqref{genkin}, it follows:
\begin{equation}
\frac{T}{T_{kin}} = \frac{T}{\Delta x (s_+ - s_-)} \mathrm{ln} \left( 1 + \frac{\lambda \tau_e}{\mu \tau_e + 1}\right) \; ,
\label{explkin2}
\end{equation}
where $\lambda$ and $\mu$ are a suitable combination of the transition rates $M_{ij}$. Note that for $\tau_e \rightarrow 0$, we have $T_{kin} \rightarrow \infty$. More interesting is the limit of low frequencies of the external drive. For $\tau_e \rightarrow \infty$, it has to be $T = T_{kin}$; therefore, it follows $\mu = \lambda w/(1-w)$, where $ w \equiv \mathrm{exp} \left[-\beta \Delta x (s_+ - s_-)\right]$. Thus, for $\tau_e \gg 1/\lambda$, we have
\begin{equation}
\frac{T}{T_{kin}} \simeq 1 - \frac{T}{\Delta x (s_+ - s_-)} \frac{(1-w)^2}{\tau_e \lambda w} \; .
\label{explkinapprox}
\end{equation}
In Fig.~\ref{fig7} we show $T^{\sigma,m}_{kin}$ as a function of $\tau_e$. In particular, since $\sigma_{ij}$ is conjugated to $\Delta f$ in the hamiltonian, the asymptotic expression in Eq.~\eqref{explkinapprox} correctly approximate the behaviour of $T^\sigma_{kin}$. 

	{\it Acknowledgements.} Simulations were run at Bari ReCaS e-Infrastructure funded by MIUR through PON Research and Competitiveness 2007–2013 Call 254 Action I. G.G. acknowledges MIUR for funding (PRIN 2017/WZFTZP, “Stochastic forecasting in complex systems”).

	\bibliographystyle{apsrev4-2}
	\bibliography{library}

\end{document}